\documentclass[final]{elsarticle}

\biboptions{sort&compress}

\usepackage{hyperref}
\usepackage{amssymb}
\usepackage{amsfonts}
\usepackage{amsmath}
\usepackage{graphicx}
\usepackage{dcolumn}
\usepackage{bm}
\usepackage{color}
\usepackage{natbib}

\newcommand{\sslash}{\mathbin{/\mkern-4mu/}}

\journal{Communications in Nonlinear Science and Numerical Simulation}

\DeclareMathOperator{\sech}{sech}
\graphicspath{{./fig-jpg/}{./fig-ps/}{./figures/}}

\begin{document}

\begin{frontmatter}

\title{Breather stripes and radial breathers of the two-dimensional sine-Gordon equation}

\author[UMass,Oxford]{P. G. Kevrekidis}
\author[SDSU]{R. Carretero-Gonz\'{a}lez\corref{cor1}}
\author[USevilla1,USevilla2]{J. Cuevas-Maraver}
\author[UAthens]{D. J. Frantzeskakis}
\author[Rouen]{J.-G. Caputo}
\author[TelAviv,Chile]{B. A. Malomed}

\cortext[cor1]{Corresponding author. Email: rcarretero@sdsu.edu}

\address[UMass]{Department of Mathematics and Statistics,
University of Massachusetts,
Amherst, MA 01003-4515, USA}

\address[Oxford]{Mathematical Institute, University of Oxford, Oxford, OX2 6GG, UK}

\address[SDSU]{Nonlinear Dynamical System Group,$^1$\fnref{NLDS}
Computational Science Research Center,$^2$\fnref{CSRC}
and Department of Mathematics and Statistics,
San Diego State University,
San Diego, CA 92182-7720, USA}

\address[USevilla1]{Grupo de F\'{\i}sica No Lineal, Universidad de Sevilla. Departamento de
F\'{\i}sica Aplicada I, Escuela Polit\'ecnica Superior,
C/ Virgen de \'{A}frica, 7, 41011 Sevilla, Spain}

\address[USevilla2]{Instituto de Matem\'aticas de la Universidad de Sevilla.
Avda. Reina Mercedes, s/n. Edificio Celestino Mutis,
41012 Sevilla, Spain}

\address[UAthens]{Department of Physics, National and Kapodistrian University of Athens,
Panepistimiopolis, Zografos, Athens 15784, Greece}

\address[Rouen]{Laboratoire de Math\'ematiques, INSA de Rouen Normandie\\ 76801 Saint-Etienne du Rouvray, France.}

\address[TelAviv]{Department of Physical Electronics, School of Electrical Engineering,
Faculty of Engineering, Tel Aviv University, Tel Aviv 69978, Israel }

\address[Chile]{Instituto de Alta Investigaci\'{o}n, Universidad de Tarapac\'{a}, Casilla
7D, Arica, Chile}

\fntext[NLD]{\texttt{URL}: \href{http://nlds.sdsu.edu}{http:$\sslash$nlds.sdsu.edu}}
\fntext[CSRC]{\texttt{URL}: \href{http://www.csrc.sdsu.edu}{http:$\sslash$www.csrc.sdsu.edu}}

\begin{abstract}
We revisit the problem of transverse instability of a 2D breather stripe of
the sine-Gordon (sG) equation. A numerically computed Floquet spectrum of
the stripe is compared to analytical predictions developed by means of
multiple-scale perturbation theory showing good agreement in the
long-wavelength limit. By means of direct simulations, it is found that the
instability leads to a breakup of the quasi-1D breather in a chain of
interacting 2D radial breathers that appear to be fairly robust in the dynamics.
The stability and dynamics of radial breathers in a finite domain are studied in
detail by means of numerical methods. Different families of such solutions
are identified. They develop small-amplitude spatially oscillating tails
(``nanoptera") through a resonance of higher-order
breather harmonics with linear modes (``phonons") belonging
to the continuous spectrum. These results demonstrate the ability of the 2D
sG model within our finite domain computations
to localize energy in long-lived, self-trapped breathing excitations.
\end{abstract}


\end{frontmatter}


\section{Introduction}

One of the most influential models in studies of solitary waves is the
sine-Gordon (sG) equation, which has been extensively explored in numerous
volumes~\cite{eilbeck,dauxois,sgbook} and reviews~\cite{kivsharmalomed}. The
one-dimensional (1D) version is the first nonlinear equation whose
integrability was found (in the form of the B\"{a}cklund transform, ca.~140
years ago~\cite{Backlund}). Later, its complete integrability was
systematically investigated by means of the inverse scattering transform
(IST)~\cite{akns}. Commonly known exact solutions to the sG equation are
topological kink solitons and kink-antikink bound states in the form of
breathers. Modes of the latter type are uncommon in continuum systems, but
find broad realizations in their discrete counterparts~\cite{flach,aubry}.

While the 1D sG equation has become a textbook etalon of integrable models,
far less is known about higher dimensional versions of the same equation,
which are not integrable by means of the IST. In particular, some effort has been
dedicated to the kinematics and dynamics of kinks in two and three
dimensions (2D and 3D)~\cite{christiansen,geicke,bogolub,samuelsen},
including their ability to produce breathers as a result of collisions with
boundaries~\cite{caputo}, and proneness to be stably pinned by local
defects~\cite{danaila}.

Nevertheless, despite the relevance of physical realizations of the sG
model in higher dimensions (see, e.g., Ref.~\cite{boris} for a recent
example), the study of breathers in such settings has been scarce. While
effective equations of motion for breathers were derived long ago~\cite{maslov},
these equations do not apply to steady isotropic breather
profiles.
Hence, the temporal dynamics of the latter states remains somewhat unclear.

Moreover, there is a relevant related question concerning quasi-1D breather
solutions in the higher-dimensional problem, i.e., breather stripes and
planes in the 2D and 3D settings, respectively. The seminal work
of Ref.~\cite{kodama} had predicted that such quasi-1D solutions are subject to
transverse instabilities. This is a natural property due to the relation of
the sG model, in the limit when it produces broad small-amplitude breathers,
to the nonlinear Schr{\"{o}}dinger (NLS) equation~\cite{eilbeck,dauxois}, in
which both bright and dark quasi-1D solitons are prone to transverse
instabilities~\cite{kidep} ---see, e.g., theoretical work of Ref.~\cite{kuzne} and
Refs.~\cite{watching,tikh,decon} for experimental realizations in atomic and
optical physics. On the other hand, quasi-1D sG kinks are \textit{not}
vulnerable to the transverse instability. In this work, we aim to further
explore the dynamics of quasi-1D sG breathers in the 2D geometry, including
their transverse instability and its development into radially shaped
breathing modes.

We start the analysis by revisiting the key result of Ref.~\cite{kodama} for
quasi-1D breathers in the 2D sG equation. We test this result for breathers
with different frequencies $\omega_{\mathrm{b}}$ and demonstrate that
their instability in the long-wavelength limit is
accurately captured by the analytical treatment. Our numerical
calculation of the respective Floquet multipliers (FMs) additionally
permits the systematic identification of spectra of transverse-instability modes
as a function of the corresponding perturbation wavenumber, $k_{y}$. These
results indicate that the quasi-1D sG breathers do not only maintain a main
band of unstable wavenumbers, similarly to what is known for the NLS
equation, where only a band of low wavenumbers is
unstable. Importantly, they also
display, for appropriate frequencies $\omega_{\mathrm{b}}$, instability
bubbles for higher wavenumbers. Near the NLS limit of $\omega_{\mathrm{b}}\rightarrow 1$,
we demonstrate how the NLS reduction, as well as a novel,
to the best of our knowledge, variational approximation involving the
transverse degree of freedom are able to capture instability and dynamical
features, including the basic necking phenomenon occurring in the course of
the breather stripe evolution.

Studying the dynamics ensuing from the generic long-wavelength instabilities
of the breather stripes, we identify spontaneous formation of localized
breather waveforms. Further, ``naively" initializing a
quasi-1D radial breather profile, with $x$ replaced by the radial variable,
$r$, we find that the solutions adjust themselves and their frequencies in
the course of a transient stage of the evolution, to form potentially very
long-lived localized modes trapping the energy in a localized spatial region. This
finding motivates us to study the radial time-periodic breathers, which we are
able to find, as genuine periodic states, using fixed-point iteration
techniques. Stability of these radial breathers is studied numerically in
the framework of the Floquet theory.

The manuscript is organized as follows. Section~\ref{sec:stripes} reports
the results concerning the breather stripes. In particular we expand the
instability results for the stripes from Ref.~\cite{kodama}. The results,
which are valid from the long-wavelength limit to larger wavenumbers, are
obtained by connecting the sG dynamics to the NLS limit, using
multiple-scale expansions, and also by dint of a variational approach, which
is relevant for large $\omega_{\mathrm{b}}$. We present numerical results
demonstrating that the development of the breather-stripe instability
nucleates long-lived radial breathers. The existence and stability of such
radial breathers is the subject of Sec.~\ref{sec:radial}. Finally, in
Sec.~\ref{sec:conclu} we present a summary of the results, concluding remarks,
and point out avenues for further research.

\section{Breather stripes}
\label{sec:stripes}

\subsection{Modulational instability of breather stripes}

The sG equation is a special case of the more general Klein-Gordon class of
models which in 2D is written as
\begin{equation}
u_{tt}-\nabla^{2}u+V^{\prime }(u)=0,
\end{equation}
where $u(x,y,t)$ is a real field depending on coordinates $(x,y)$ and time $t$,
subscripts with respect to the independent variables denote partial
derivatives, and $V(u)$ is the potential that defines the particular model.
The most studied among these models pertain to $V(u)=1-\cos (u)$ and
$V(u)=(u^{2}-1)^{2}/2$, which are referred to as the sG~\cite{sgbook} and
``$\phi^{4}$"~\cite{ourphi4} equations, respectively. From now on, we focus on the former one:
\begin{equation}
u_{tt}-\nabla^{2}u+\sin u=0.
\label{eq:sG}
\end{equation}
The 1D version of the sG equation admits commonly known kink/antikink and
breather solutions. Kinks are (generally speaking, traveling) 1D profiles
that asymptote to the uniform background states, $u_{\pm \infty }=\{0,2\pi\}$,
of the form
\begin{equation}
u_{\mathrm{k}}=4\arctan \left[ \exp \left( s\left( x-ct\right)
/\sqrt{1-c^{2}}\right) \right] ,  \label{kink}
\end{equation}
where $s=\pm 1$ is the topological charge, which distinguishes kinks and
antikinks, and $c$ is the velocity, which may take values $|c|<1$.
On the other hand, breathers correspond to profiles which are oscillatory in
time and localized in space. In particular, the exact 1D sG breather solution is
\begin{equation}
u_{\mathrm{b}}(x,t)=4\tan^{-1}\!\left[ {\frac{\beta }{\omega_{\mathrm{b}}}}
\,{\sech}\left( \beta \,x\right) \cos \left( \omega_{\mathrm{b}}\,t\right)
\right] ,  \label{eq:1D_breather}
\end{equation}
where $\beta \equiv \sqrt{1-\omega_{\mathrm{b}}^{2}}$ and the band of the
breather frequencies is $0\leq \omega_{\mathrm{b}} < 1$.

In the spirit of the work done in Ref.~\cite{danaila} for a kink stripe,
we consider the transverse instability of the breather stripe
\begin{equation}
u_{\mathrm{2D}}(x,y,t)=u_{\rm b}(x,t).
\end{equation}
We then perturb the breather stripe according to
%
$$ u(x,y,t)= u_{\rm b}(x,t) + w(x,y,t)$$
where the perturbation $w$ is assumed small.
Enforcing this solution to satisfy Eq.~(\ref{eq:sG}) yields, to first order,
the following evolution equation for the perturbation:
\begin{equation}
w_{tt} - (w_{xx} + w_{yy}) + \cos(u_{\rm b}(x,t)) w =0 .
\label{wtt}
\end{equation}
This equation is a linear, inhomogeneous in $x$, wave equation
so that we can study each transverse wave number $k_y$ separately and assume
\begin{equation}
w (x,y,t)=\xi (x,t)\exp (ik_{y}y),
\label{wky}
\end{equation}
where $\xi(x,t)$ describes the $x$-dependent shape (eigenmode) of the perturbation.
Plugging (\ref{wky}) into (\ref{wtt}), yields
\begin{equation}
\xi_{tt} - \xi_{xx} + \left ( k_{y}^2 + \cos(u_{\rm b}(x,t) \right ) \xi =0.
\label{xitt}
\end{equation}
This is a periodically forced wave equation and can be solved using
Floquet theory, see Appendix for details. The
final result is the linear operator ${\cal M}$
\begin{equation}
\left(
\begin{array}{c}
\xi(x,T) \\
\xi_{t}(x,T) \\
\end{array}%
\right) =\mathcal{M}\left(
\begin{array}{c}
\xi(x,0) \\
\xi_t(x,0) \\
\end{array}%
\right) ,  \label{monodromyc}
\end{equation}
relating the solutions of Eq.~(\ref{xitt}) at $t=0$ and $t=T$.
If the maximal Floquet multiplier $\Lambda $ (obtained
from the eigenvalues of $\mathcal{M}$) is such that
$|\Lambda|> 1$, $\xi$ will grow and the stripe will be unstable.
In the case of instability (where at least one eigenvalue has modulus larger than 1)
the eigenmode $\psi$ corresponding to the maximum  $|\Lambda|$
gives the solution $\xi(x,t)$ and the modulation of the stripe
through its perturbation $w(x,y,t)$.

In the seminal work of Ref.~\cite{kodama}, Ablowitz and Kodama (AK) predicted
that, in the long-wavelength limit, breather stripes
are subject to  modulational instability (MI).
Their main result provides the instability
growth rate $\lambda$ for the stripe, as a function of the perturbation
wavenumber, $k_{y}$:
\begin{equation}
\lambda =\sigma k_{y},\quad \sigma ^{2}={\frac{\sqrt{1-\omega_{\mathrm{b}}^{2}}}
{\omega_{\mathrm{b}}}\arcsin \left( \sqrt{1-\omega_{\mathrm{b}}^{2}}\right) }.
\label{eq:AK}
\end{equation}

Given that the AK result is valid in the $k_{y}\rightarrow 0$ limit,
we wish to investigate the instability for larger values of $k_{y}$.
To this end, one needs to solve Eq.~(\ref{eq:sG})
numerically, by means of a suitable fixed-point method (see details in the Appendix),
taking into regard that a finite size (and a finite discretization parameter) of the
solution domain induces gaps in the spectrum of the linear modes and,
also that breathers may feature modified tails generated by
``hybridization" (mixing) of the third and higher-order odd harmonics with
linear modes (``phonons"), reminiscent of the formation of the so-called
``nanoptera" in the $\phi^{4}$ equation~\cite{sgbook}. The
hybridization depends on the location of the linear modes, which, in turn,
depends on the specifics of the domain size and boundary conditions. We have
used periodic boundary conditions in domain involving $x\in \lbrack -L_{x},L_{x})$
with $L_{x}=100$ and computed the periodic steady states corresponding
to zero initial velocity $u_t(x,y,t=0)=0$.
Finally, it is important to note that, addressing the dynamics with the
underlying time-periodic solution, one needs to resort to Floquet
analysis and the computation of FMs, in order to
extract the (in)stability eigenvalues (see details in Appendix). The FMs,
$\Lambda$, are translated into (in)stability eigenvalues $\lambda$ (in
particular, for the comparison to the AK theory), via $\Lambda =\exp(\lambda T)$,
where $T=2\pi/\omega_{\rm b}$ is the breather period.

\begin{figure}[htb]
\begin{center}
\includegraphics[width=8.0cm]{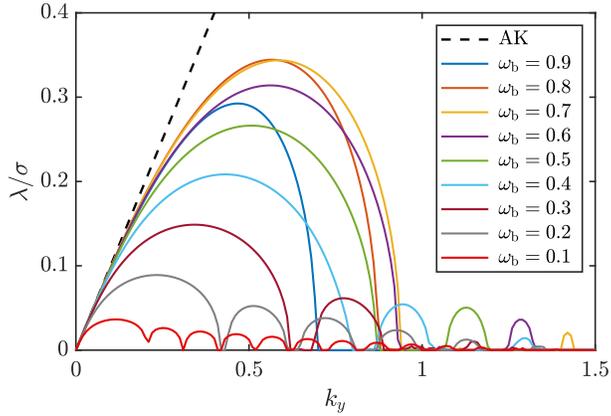}
\end{center}
\caption{(Color online) Modulational-instability eigenvalues for the
breather stripe vs.~wavenumber $k_{y}$ of the modulational
perturbations, for the indicated values of the breather frequency. Solid
curves depict the eigenvalues numerically determined by unstable Floquet
multipliers (see the text). The dashed line represents the analytical
Ablowitz-Kodama (AK) result for the long-wavelength limit, $k_{y}\rightarrow 0$
\cite{kodama}, given by Eq.~(\ref{eq:AK}).}
\label{fig:eigen_stripe}
\end{figure}

\begin{figure}[htb]
\begin{center}
\includegraphics[width=8.0cm]{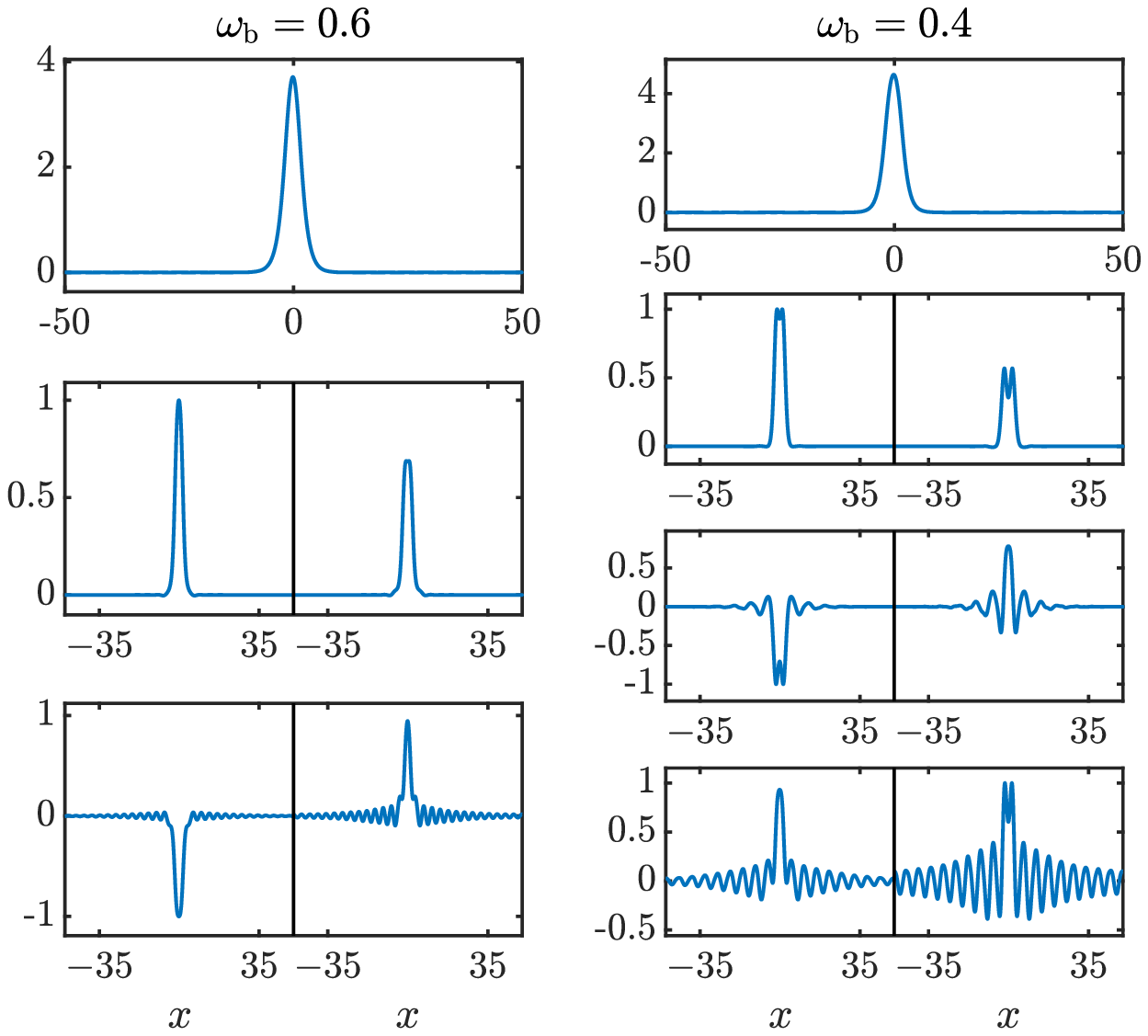}
\end{center}
\caption{(Color online)
Snapshots of the breathers $u_{\rm b}(x,t=0)$
and corresponding eigenfunctions $\psi$
for different MI windows/bubbles. The left and right sets of panels
correspond to $\omega_{\mathrm{b}}=0.6$ and
$\omega_{\mathrm{b}}=0.4$, respectively.
Snapshots of the breathers $u_{\rm b}(x,t=0)$ are displayed in the top
panels. The subsequent ones, from top to bottom, display the $u$ (left)
and $u_{t}$ (right) components of the eigenfunctions $\psi$
corresponding to the highest-MI wavenumbers in the bubbles of
Fig.~\ref{fig:eigen_stripe}. For $\omega_{\mathrm{b}}=0.6$, two MI bubbles exist
(with maxima located at $k_y\approx 0.57$ and 1.28), while for
$\omega_{\mathrm{b}}=0.4$ there are three
MI bubbles (with maxima located at $k_y\approx 0.43$, $0.94$ and 1.30).
The computation domain is $x\in \lbrack -100,100)$, half of which is
shown here.
}
\label{fig:evecs}
\end{figure}

Figure~\ref{fig:eigen_stripe} depicts the dependence on $k_{y}$ of the MI growth
rate for selected values of the breather stripe frequency, $\omega_{\mathrm{b}}$.
The eigenvalues have been rescaled by $\sigma$ [see Eq.~(\ref{eq:AK})]
to facilitate the comparison to the AK analytical prediction,
given by Eq.~(\ref{eq:AK}) and shown by the dashed line in the figure. The
MI spectrum indeed follows the AK prediction as $k_{y}\rightarrow 0$, for
all values of $\omega_{\mathrm{b}}$, i.e., for long perturbation
wavelengths, the MI eigenvalue is $\lambda/\sigma \approx  k_{y}$,
independent of $\omega_{\mathrm{b}}$. Further, Fig.~\ref{fig:eigen_stripe}
demonstrates that, beyond the AK limit, $\lambda$ grows with $k_{y}$ until
reaching a maximum, and then decreases, falling to zero at some
$k_{y}=\tilde{k}_{y}$. We refer to this first instability window ($0<k_{y}<\tilde{k}_{y}$)
as the main MI ``bubble''.
The value $\tilde{k}_{y}$, as well as the maximum of
$\lambda/\sigma $, increase with $\omega_{\mathrm{b}}$ for $\omega_{\mathrm{b}}\leq 0.7$,
but this trend changes to a decrease for $0.7<\omega_{\mathrm{b}}<1$.
This result, which is in stark contrast with properties of MI for the
NLS equation~\cite{Cisneros,JiankeBook:2010}, has its origin in increasing
contributions from the third and higher harmonics of the breather's
oscillatory waveform when $\sqrt{1-\omega_{\mathrm{b}}^{2}}$ decreases
and, consequently, truncating at the first harmonic is no longer a good
approximation for expression (\ref{eq:1D_breather}) for $\omega_{\mathrm{b}}\lesssim 0.7$.

Furthermore, beyond the main MI bubble, Fig.~\ref{fig:eigen_stripe} suggests
another noteworthy feature: the existence of secondary MI bubbles for $k_{y}>\tilde{k}_{y}$.
These secondary bubbles get narrower and more spaced with the
increase of $\omega_{\mathrm{b}}$; in fact, they are not found for
$\omega_{\mathrm{b}}>0.7$. To understand MI in the secondary bubbles,
and to compare it to the main MI window, $k_{y}<\tilde{k}_{y}$, we explore the shape of the
corresponding eigenfunctions of the modulational perturbations. In
particular, Fig.~\ref{fig:evecs} depicts the breather profile and eigenfunctions
corresponding to unstable eigenvalues in the main MI window and secondary
bubbles for $\omega_{\mathrm{b}}=0.6$ and $\omega_{\mathrm{b}}=0.4$. In
the case of $\omega_{\mathrm{b}}=0.6$ (the set of left panels in the
figure), MI takes place in the main window and in one secondary bubble, while in
the case of $\omega_{\mathrm{b}}=0.4$ (the set of right panels) two
secondary MI bubbles are present. All perturbation eigenfunctions in these
MI bubbles are real as the corresponding eigenvalue $\Lambda$ responsible for the
instability is also real (we depict the respective components $u$ and $u_{t}$ in
the left and right plots, respectively). Figure~\ref{fig:evecs} exhibits
eigenfunctions that bear qualitatively similar shapes. Namely, the
eigenfunctions in the main MI window are all localized without oscillating
tails, while in the secondary bubbles, tails are attached to the core of the
eigenfunctions, getting stronger in higher-order bubbles.

\subsection{Connection to NLS}

To better understand the MI spectrum for the breather stripes, we may use
the connection between the sG and the NLS equations, in the limit
of $\omega_{\mathrm{b}}\rightarrow 1$.
Such a connection can be accurately established
via the multiple-scales perturbation method~\cite{jeffrey}. Below we will
briefly present the methodology and provide the main results (see also
Refs.~\cite{eilbeck,dauxois} for a detailed analysis). First we note that, in the
limit of $\omega_{\mathrm{b}}\rightarrow 1$, the sG breather of Eq.~(\ref{eq:1D_breather})
has a small amplitude and, hence, one can expand the sG
nonlinearity as $\sin u\approx u-u^{3}/6$. Thus, we seek solutions of
the resulting $\phi^{4}$-equation in the form of the asymptotic expansion:
\begin{equation}
u(x,t)=\sum_{n=1}^{N}\epsilon^{n}u_{n}(x_{0},\ldots ,x_{N},~t_{0},\ldots,t_{N}),
\label{asex1}
\end{equation}
where $u_{n}$ are functions of the independent variables $x_{j}=\epsilon^{j} x$
and $t_{j}=\epsilon^{j}t$ ($j=0,1,2,\ldots N$), while $0<\epsilon \ll 1$ is
a formal small parameter. Substituting ansatz~(\ref{asex1}) in the equation,
and collecting coefficients in front of different powers of $\epsilon$
yields a set of equations,
the first three ones being
\begin{eqnarray}
O(\epsilon^{1}) &:&\mathcal{L}_{0}u_{1}=0,  \label{o1} \\
O(\epsilon^{2}) &:&\mathcal{L}_{0}u_{2}+\mathcal{L}_{1}u_{1}=0,  \label{o2}
\\
O(\epsilon^{3}) &:&\mathcal{L}_{0}u_{3}+\mathcal{L}_{1}u_{2}+\mathcal{L}_{2}u_{1}
-\frac{1}{6}u_{1}^{3} = 0.  \label{o3}
\end{eqnarray}
where
\begin{eqnarray}
\mathcal{L}_{0}&\equiv& \partial_{t_{0}}^{2}-\partial_{x_{0}}^{2}+1,
\notag
\\[1.0ex]
\notag
\mathcal{L}_{1}&\equiv& 2(\partial_{t_{0}}\partial_{t_{1}}-\partial_{x_{0}}
\partial_{x_{1}}),
\\[1.0ex]
\notag
\mathcal{L}_{2}&\equiv& \partial_{t_{1}}^{2}
-\partial_{x_{1}}^{2}+2\left( \partial_{t_{0}}\partial_{t_{2}}
-\partial_{x_{0}}\partial_{x_{2}}\right).
\end{eqnarray}
To leading order,
$O(\epsilon )$, the solution to Eq.~(\ref{o1}) is
\begin{equation}
u_{1}=A(x_{1},x_{2},\ldots ,t_{1},t_{2},\ldots )\,\exp (i\Phi )+\mathrm{c.c},
\label{los}
\end{equation}
where $A$ is a yet-to-be-determined function of the slow variables,
$\mathrm{c.c}$ stands for the complex conjugate, and $\Phi =kx_{0}-\omega t_{0}$,
with wavenumber $k$ and frequency $\omega $ obeying the linear dispersion
relation,
%
\begin{equation}
\omega^{2}=k^{2}+1.  \label{disp}
\end{equation}
Substituting solution~(\ref{los}) in Eq.~(\ref{o2}), it is evident that the
second term is secular and has to be removed. This yields equation
$A_{t_{1}}+v_{g}A_{x_{1}}=0$, where the $v_{g}=k/\omega $ is the sG group
velocity, as per Eq. (\ref{disp}). This result suggests that the amplitude
function $A$ depends on variables $x_{1}$ and $t_{1}$ only through a
traveling coordinate,
\begin{equation}
X=x_{1}-v_{g}t_{1},  \label{xi}
\end{equation}
i.e., $A=A(X,x_{2},x_{3},\ldots ,t_{2},t_{3},\ldots )$.
Furthermore, as concerns the solution of
Eq.~(\ref{o2}) [which is now reduced to $\mathcal{L}_{0}u_{2}=0$], we take the
trivial solution $u_{2}=0$, because a nontrivial one would be of the same form as
Eq.~(\ref{los}), hence its amplitude function can be included in field $A$.
Finally, at the next order, $O(\epsilon^{3})$, upon substituting $u_{1}$
from Eq.~(\ref{los}) and $u_{2}=0$, Eq.~(\ref{o3}) becomes:
\begin{equation}
\mathcal{L}_{0}u_{3}=\left[ \left( 1-v_{g}^{2}\right) A_{XX}+2i\omega \left(
A_{t_{2}}+v_{g}A_{x_{2}}\right) \right] \exp (i\theta )
+\frac{1}{6}A^{3}\exp (3i\theta )+\frac{1}{2}|A|^{2}A\exp (i\theta )
+\mathrm{c.c}=0.  \label{o3eq}
\end{equation}
It is observed that secular terms $\propto \exp (i\theta )$ arise on the
right-hand side of Eq.~(\ref{o3eq}) [term $\propto \exp (3i\theta )$ is not
secular, because it is out of resonance with the uniform solution]. Removing
the secular terms leads to the following NLS equation for complex amplitude $A$:
\begin{equation}
iA_{t_{2}}+\left( 2\omega^{3}\right)^{-1}A_{XX}+\left( 4\omega \right)
^{-1}|A|^{2}A=0,  \label{nlsf}
\end{equation}
where we have removed the group-velocity term by redefining $x_{2}\mapsto
x_{2}-v_{g}t_{2}$, and made use of identity $1-v_{g}^{2}=1/\omega^{2}$,
resulting from Eq. (\ref{disp}).

The fact that the NLS equation~(\ref{nlsf}) is valid in the band $\omega \geq 1$,
according to Eq.~(\ref{disp}), allows us to consider the limit of $\omega\rightarrow 1$,
i.e., $k\rightarrow 0$. In this case, with the vanishing
group velocity $v_{g}\rightarrow 0$, variable $X$, defined by Eq.~(\ref{xi}),
carries over into $x_{1}$, Eq.~(\ref{nlsf}) reduces to the form of
\begin{equation}
iA_{t_{2}}+\frac{1}{2}A_{x_{1}x_{1}}+\frac{1}{4}|A|^{2}A=0.  \label{NLS2}
\end{equation}
The stationary soliton solution of the reduced NLS, expressed in terms of
original variables $x$ and $t$, is
\begin{equation}
A=\eta\,\sech[(1/2)(\epsilon \eta x)]\exp [(i/8)\epsilon^{2}\eta^{2}t],
\label{soliton}
\end{equation}
where $\eta$ is an arbitrary $O(1)$ parameter. Thus, an approximate sG
breather solution, valid up to order $O(\epsilon)$ near the edge of the
phonon band, is produced by Eqs.~(\ref{los}) and (\ref{soliton}):
\begin{equation}
u(x,t)\approx 2\epsilon \eta ~\sech\left( \frac{1}{2}\epsilon \eta
x\right) \cos \left[ \left( 1-\frac{1}{8}\epsilon^{2}\eta^{2}\right)
t \right] .
\label{apsol}
\end{equation}
Note that the frequency of the soliton solution~(\ref{apsol}),
$\omega_{\mathrm{s}}=1-(\epsilon^{2}\eta^{2}/8)$,
is \textit{smaller} than the cutoff
frequency, $\omega =1$, which naturally means that this self-trapped
excitation belongs to the phonon \textit{bandgap}, as is known for
the sG breather. This fact indicates a strong connection between the
stationary NLS soliton and the sG breather in the limit of
\begin{equation}
0<1-\omega_{\mathrm{b}}\ll 1.
\label{eq:01}
\end{equation}
Indeed, in this limit, observing that the argument of $\tan^{-1}$ in
solution~(\ref{eq:1D_breather}) is small, it may be approximated by
$u_{\mathrm{b}}\approx (4\beta /\omega_{\mathrm{b}})\sech(\beta x)\cos
(\omega_{\mathrm{b}}t)$. Then, letting $\epsilon \eta \equiv 2\beta$,
which means that $\omega_{\mathrm{b}}\approx 1-(\epsilon^{2}\eta
^{2}/8)\equiv \omega_{\mathrm{s}}$, we see that the approximate NLS soliton
solution, given by Eq.~(\ref{apsol}), is identical to the sG breather in the
limit case defined by Eq.~(\ref{eq:01}). Accordingly, it is expected that the
breather stripe dynamics can be adequately approximated by the NLS
equation~(\ref{nlsf}), where $\omega$ is identified as the
breather stripe frequency, $\omega_{\mathrm{b}}$.

\begin{figure}[tbp]
\begin{center}
\includegraphics[width=8.0cm]{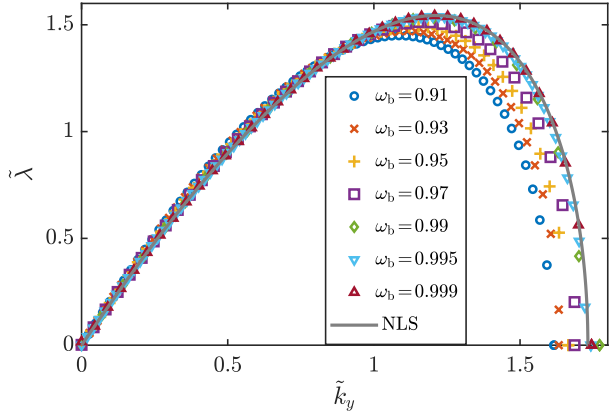}
\end{center}
\caption{(Color online) Comparison between the (scaled) instability spectrum
of the sG breather stripe (shown by colored dots) and the instability spectrum
for a stationary NLS bright-soliton stripe (shown by the gray solid line,
as per Refs.~\cite{Cisneros,JiankeBook:2010}). The wavenumber and
instability growth rate are rescaled as:
$\tilde{k}_{y}\equiv k_{y}/\sqrt{1-\omega_{\mathrm{b}}^{2}}$
and $\tilde{\lambda} \equiv 2\lambda /(1-\omega_{\mathrm{b}}^{2})$.
}
\label{fig:sG_vs_NLS}
\end{figure}

In particular, the connection between the NLS equation and the sG stripe
allows us to approximate the stability spectrum of the sG breather stripe by
that for the NLS bright-soliton stripes, which was studied in detail
previously~\cite{Kuznetsov,JiankeBook:2010,Cisneros}. Specifically,
in Fig.~\ref{fig:sG_vs_NLS} we compare the rescaled numerically computed instability
spectra of the sG breather stripe (dots) for different breather frequencies
from region~(\ref{eq:01}) with those for a stationary NLS bright-soliton
stripe, taken from Refs.~\cite{Cisneros,JiankeBook:2010}. It is evident
that, in the limit of $\omega_{\mathrm{b}}\rightarrow 1$, the (scaled)
instability spectrum of the sG stripe indeed smoothly approaches its NLS counterpart.

\subsection{Variational approach for large frequencies}
\label{sec:stripes:VA}

In this Section we present a study of the evolution of sG breather stripes,
based on the variational approximation, which is valid for relatively large
breather frequencies. To this end, we note that the 2D sG equation~(\ref{eq:sG})
can be derived from the Lagrangian, $L=\iint \mathcal{L}\,dx\,dy$, with density
\begin{equation}
\mathcal{L}=\frac{1}{2}\left( u_{t}^{2}-u_{x}^{2}-u_{y}^{2}\right)
-2\sin^{2}\left( \frac{u}{2}\right).
\label{eq:L}
\end{equation}
We rewrite the exact 1D sG breather solution (\ref{eq:1D_breather}) as
\begin{equation}
u_{\mathrm{b}}=4\tan^{-1}\left[ \left( \tan \mu \right) \,{\sin \left(
t\cos \mu \right) }\,{\sech\left( x\sin \mu \right) }\right],
\label{breather}
\end{equation}
where parameter $\mu \equiv \arcsin \left( \sqrt{1-\omega_{\mathrm{b}}^{2}}\right)$
takes values $0\leq \mu \leq \pi /2$. We now focus on broad
low-amplitude breathers in the limit of $\mu \ll \pi /2$, approximating the
exact 1D breather (\ref{breather}) by the following \textit{ansatz}:
\begin{equation}
u_{\mathrm{b}}={\chi (y,t)}\,{\sech\left( \mu x\right) }.
\label{ans}
\end{equation}
Here, $\chi$ is a variational parameter accounting for the possibility of
variations of the width of the stripe in the $y$ direction, which may
represent the so-called necking instability~\cite{Kuznetsov,gorza2}.
According to the VA, evaluating the Lagrangian with ansatz~(\ref{ans})
substituted in Eq.~(\ref{eq:L}), and deriving the Euler-Lagrange equation
from it~\cite{Progress} should yield an effective evolution equation for the
width parameter $\chi\left( y,t\right) $. In the present form, the Lagrangian
density (\ref{eq:L}) cannot be integrated analytically for ansatz~(\ref{ans}).
Nonetheless, focusing on small-amplitude breather stripes, one may expand
the nonlinear term to simplify the Lagrangian density:
\begin{equation}
\mathcal{L}=\frac{1}{2}\left( u_{t}^{2}-u_{x}^{2}-u_{y}^{2}\right)
-\frac{1}{2}u^{2}+\frac{1}{24}u^{4}.
\label{LL}
\end{equation}
Upon substituting ansatz~(\ref{ans}) in Eq.~(\ref{LL}), the integration can
be performed in the $x$ direction, yielding the effective Lagrangian:
\begin{equation}
\mu L_{\mathrm{eff}}=\int_{-\infty }^{+\infty }\left( \chi_{t}^{2}
-\chi_{y}^{2}+\left( 1-\frac{\mu^{2}}{3}\right) \chi^{2}+\frac{1}{18}\chi^{4}\right) dy.
\label{LLL}
\end{equation}
Finally, the variational equation for $\chi(y,t)$, which governs the
evolution of necking perturbations of the sG stripe under the condition~(\ref{eq:01}),
is immediately derived from the Lagrangian~(\ref{LLL}):
\begin{equation}
\chi_{tt}-\chi_{yy}+\left( 1-\frac{\mu^{2}}{3}\right) \chi -\frac{1}{9}
\chi^{3}=0.
\label{eq:chi}
\end{equation}

\begin{figure}[htb]
\begin{center}
\includegraphics[width=8.0cm]{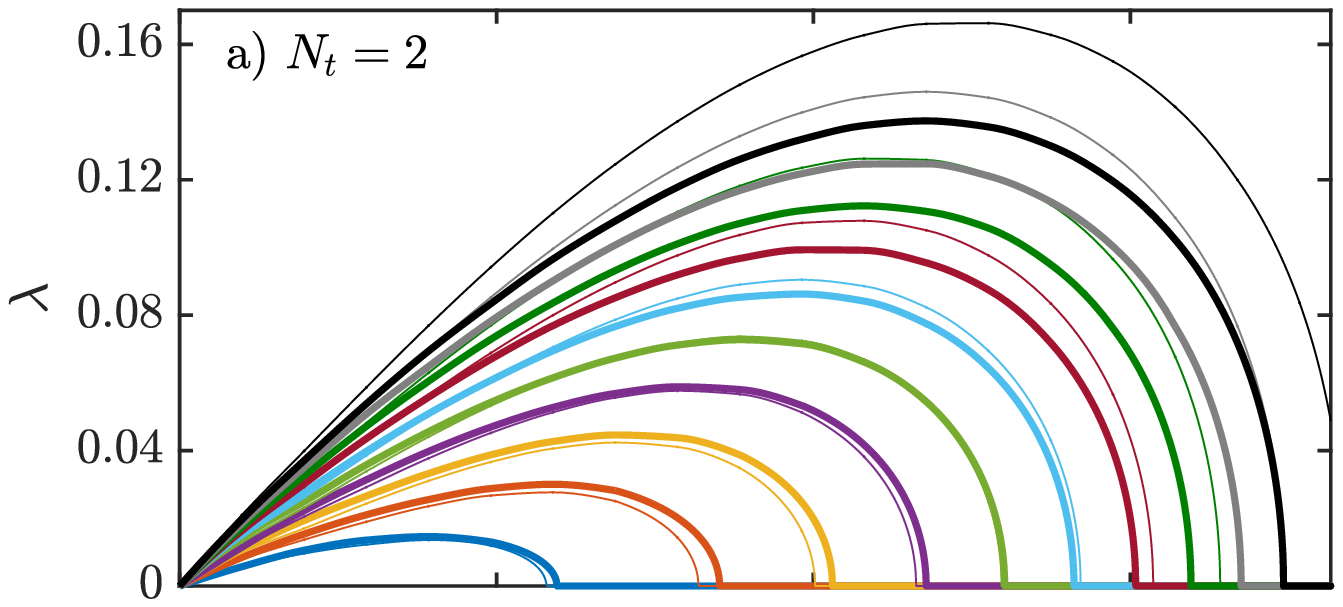}
\\[0.2ex]
\includegraphics[width=8.0cm]{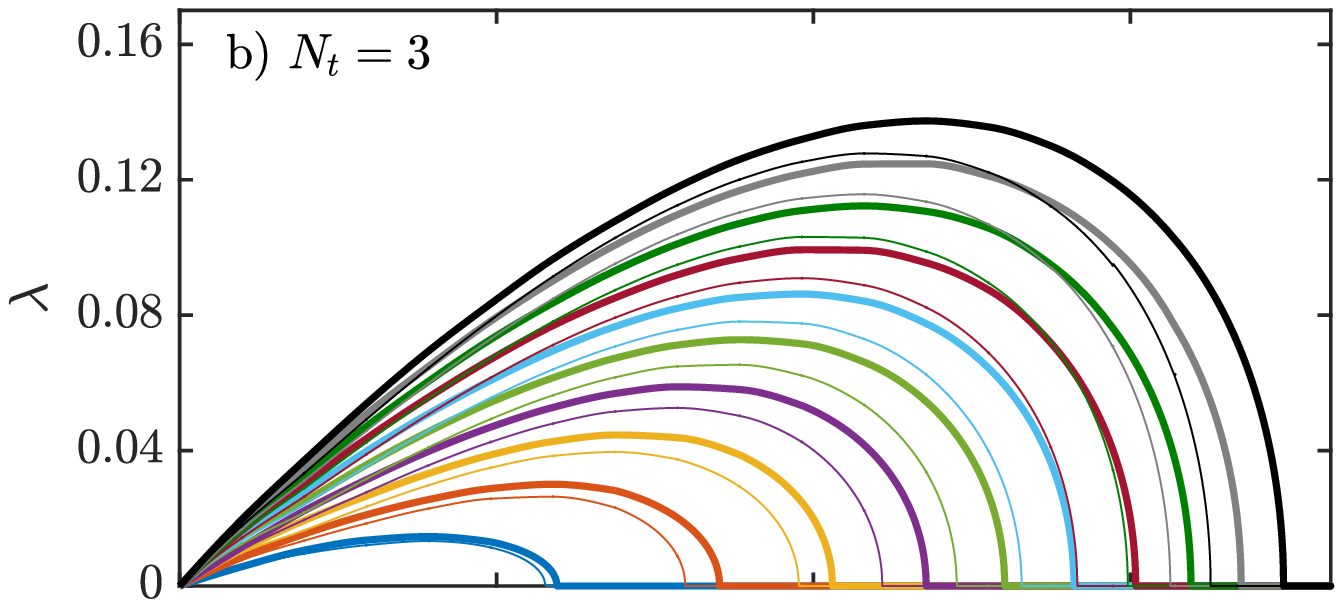}
\\[0.2ex]
\includegraphics[width=8.0cm]{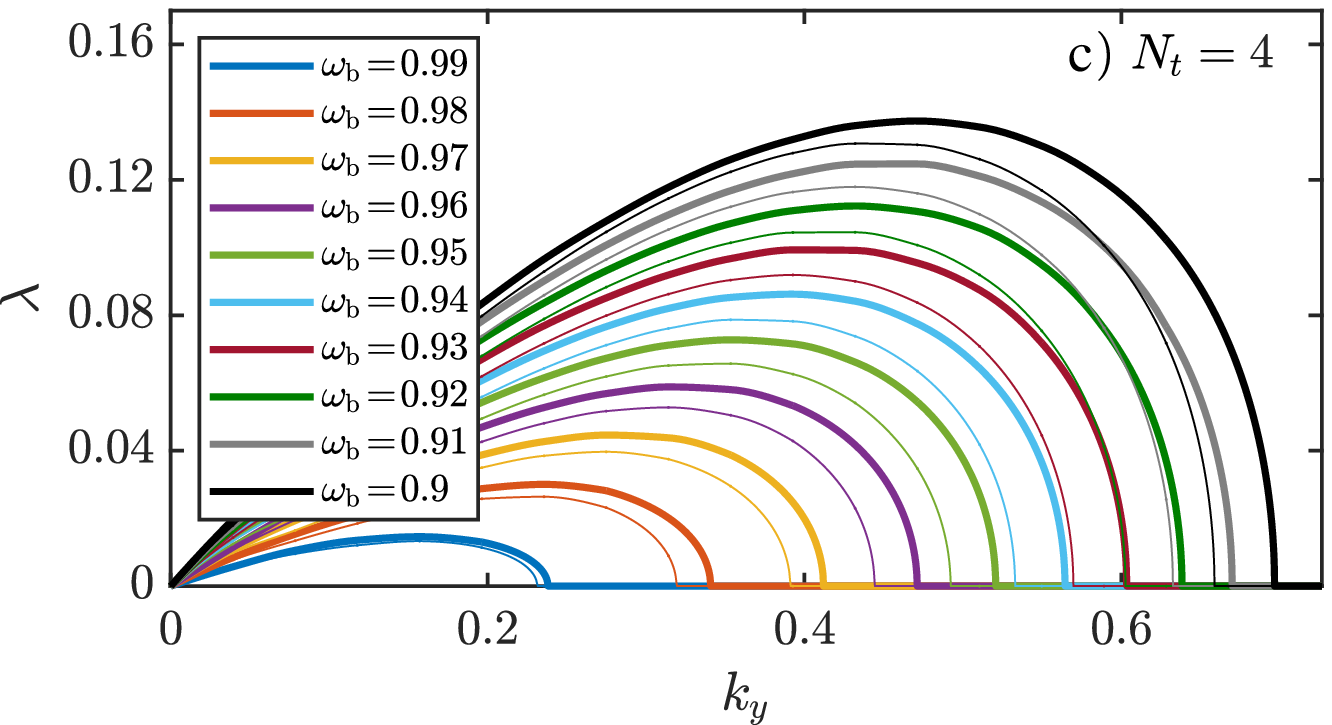}
\end{center}
\caption{(Color online) Stability spectra of the sG breather (thick lines)
and as produced by reduced model~(\ref{eq:chi_sum}) (thin lines) for
different breather frequencies. The top panel depicts the spectrum
produced by Eq.~(\ref{eq:chi_sum}) with only $N_{t}=2$ [i.e., by
Eq.~(\ref{eq:chi})], instead of the infinite sum in the full sG
equation. The second and third panels display results produced by
Eq.~(\ref{eq:chi_sum}) with $N_{t}=3$ and $4$, respectively.
There is no discernible difference in the spectra for $N_t \geq 4$.}
\label{fig:sG_vs_VA}
\end{figure}

\begin{figure*}[tbh]
\begin{center}
\includegraphics[width=4.0cm]{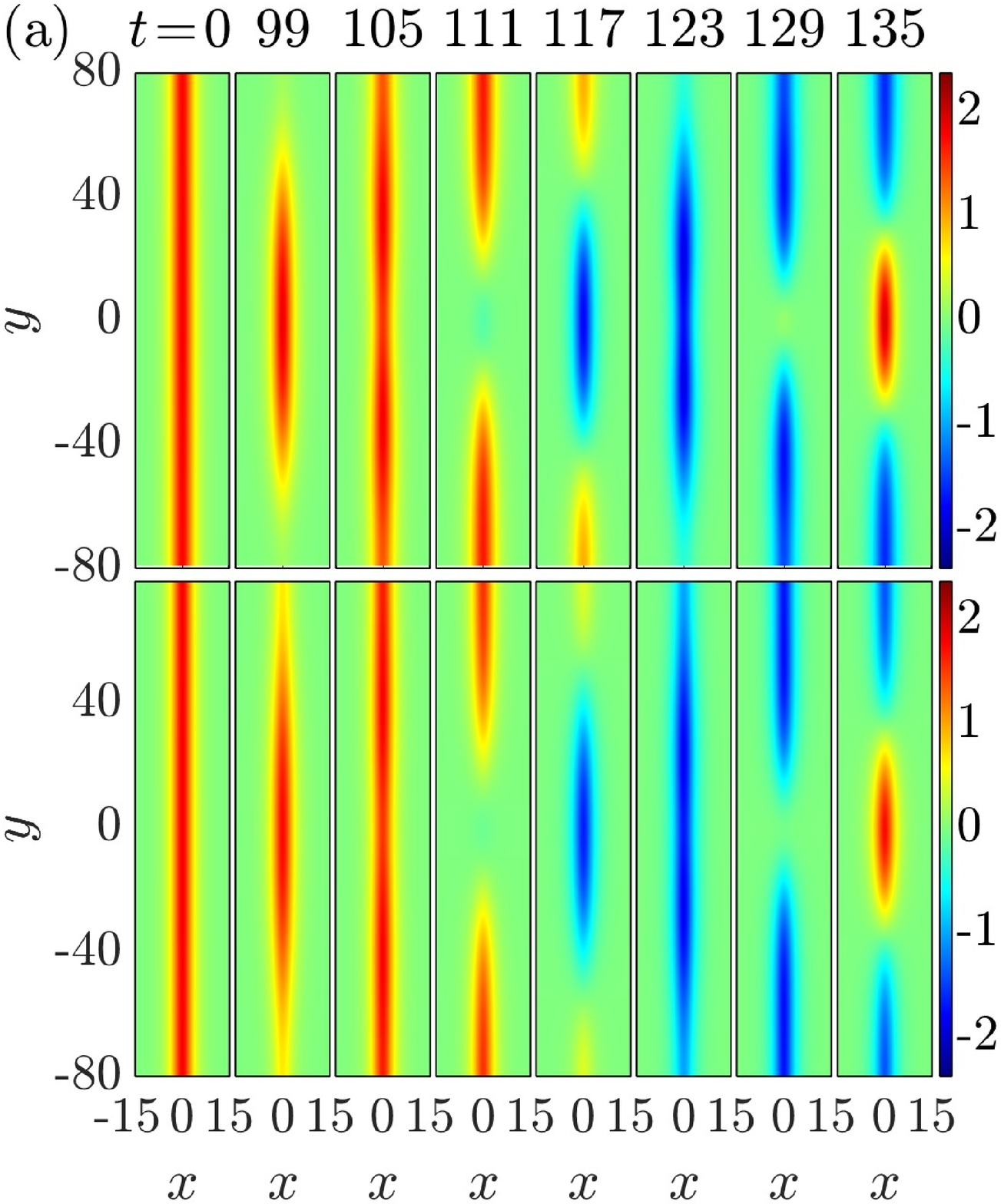}
\includegraphics[width=4.0cm]{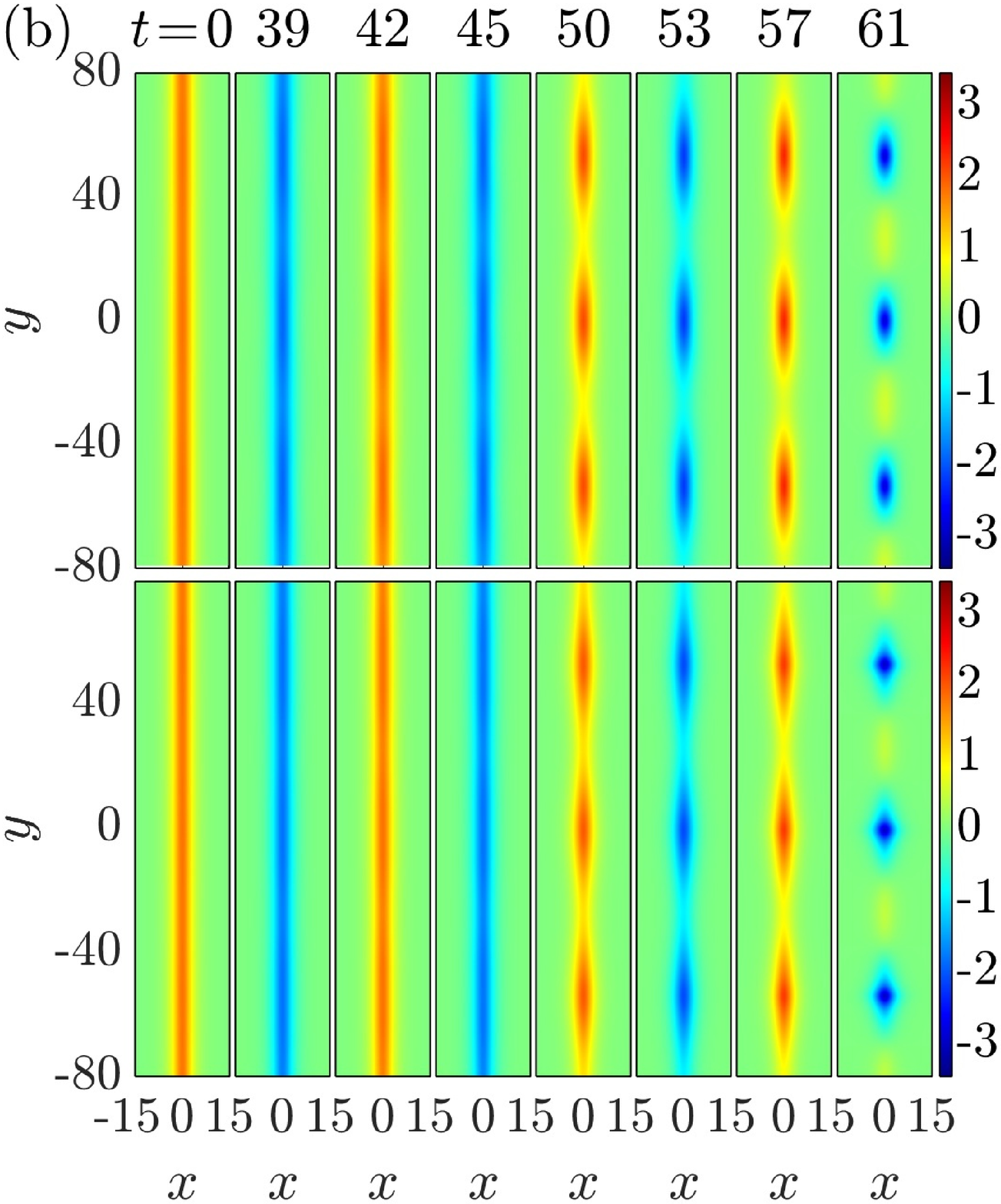}
\includegraphics[width=4.0cm]{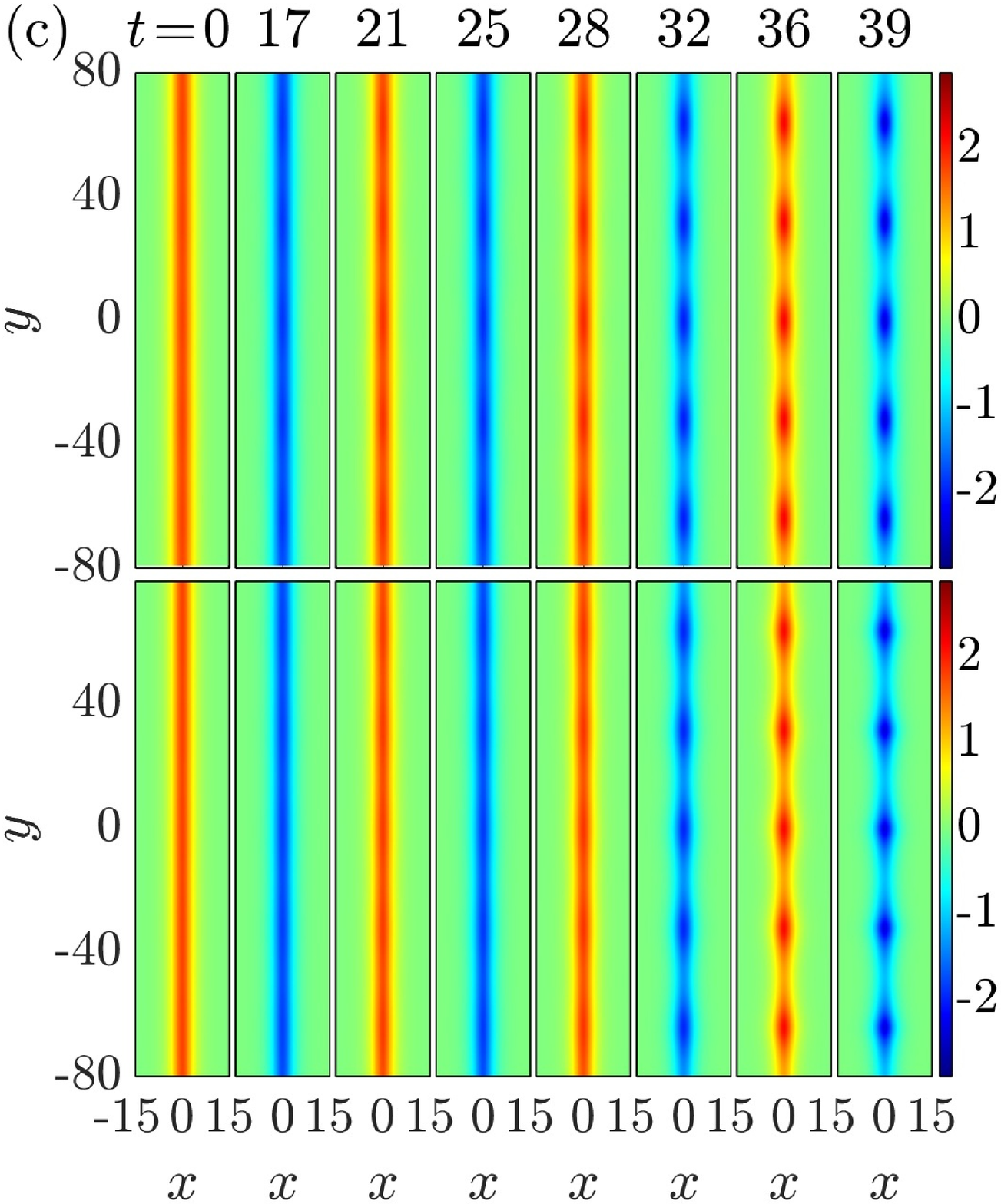}
\includegraphics[width=4.0cm]{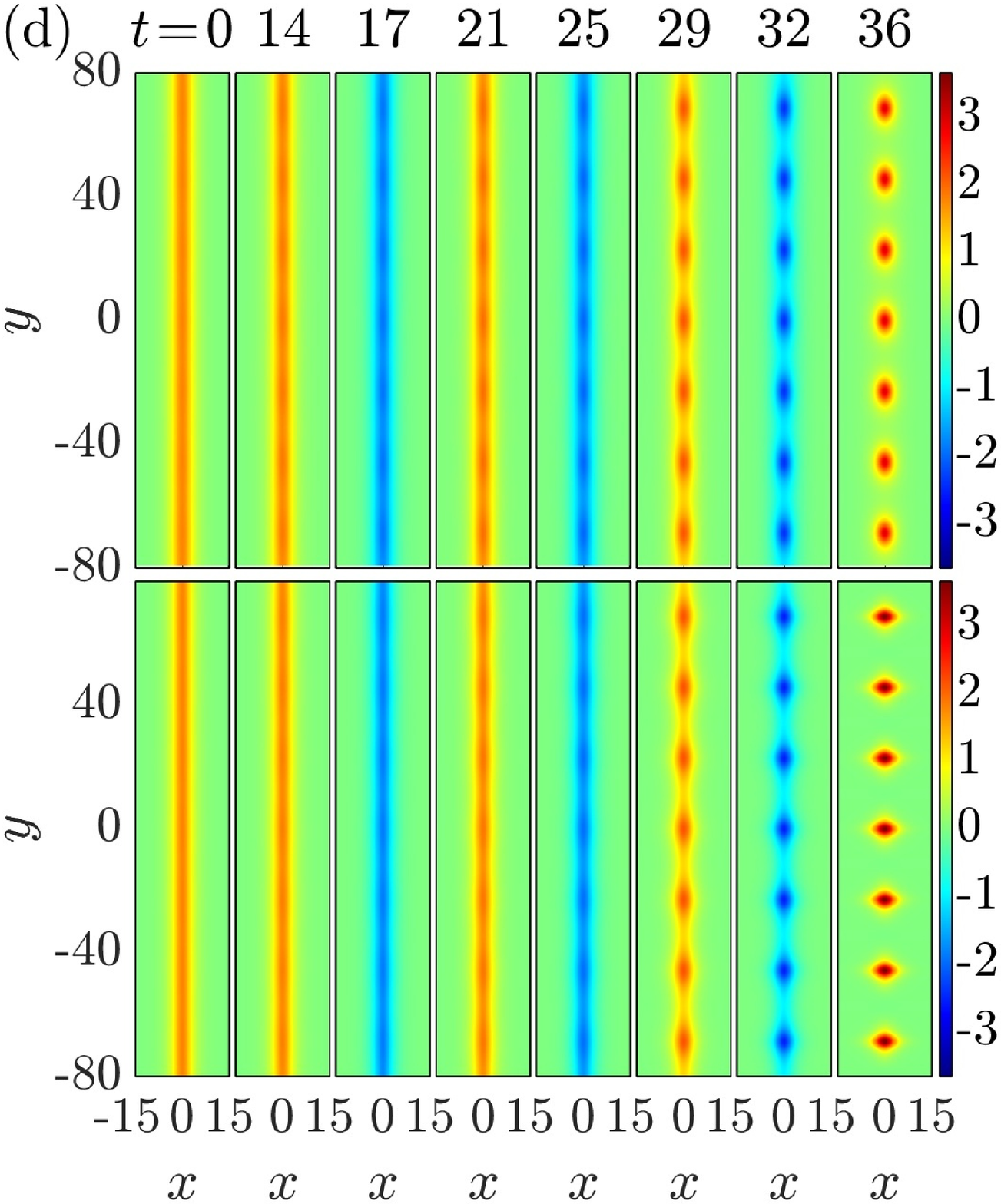}
\\[3.0ex]
\includegraphics[width=4.0cm]{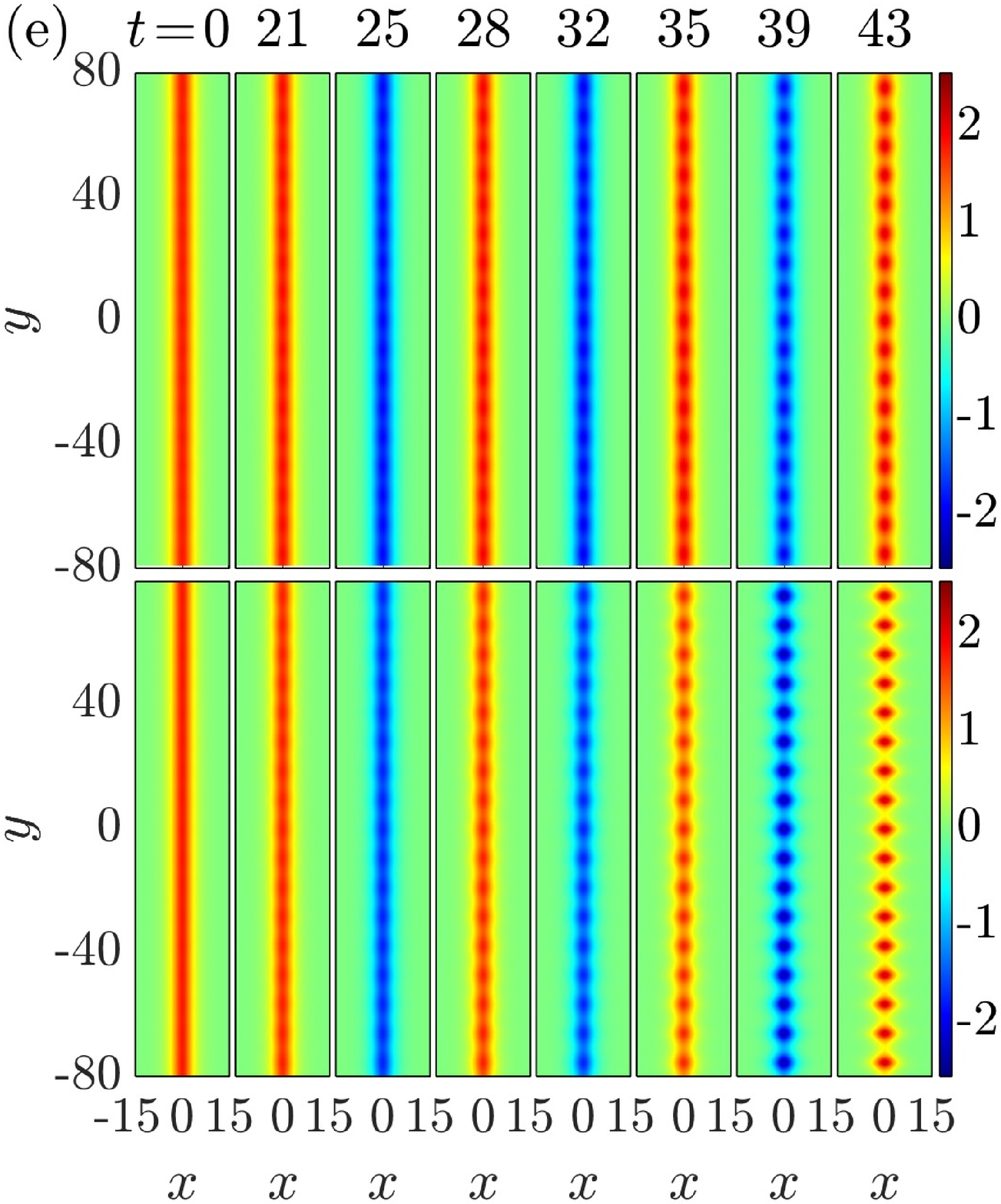}
\includegraphics[width=4.0cm]{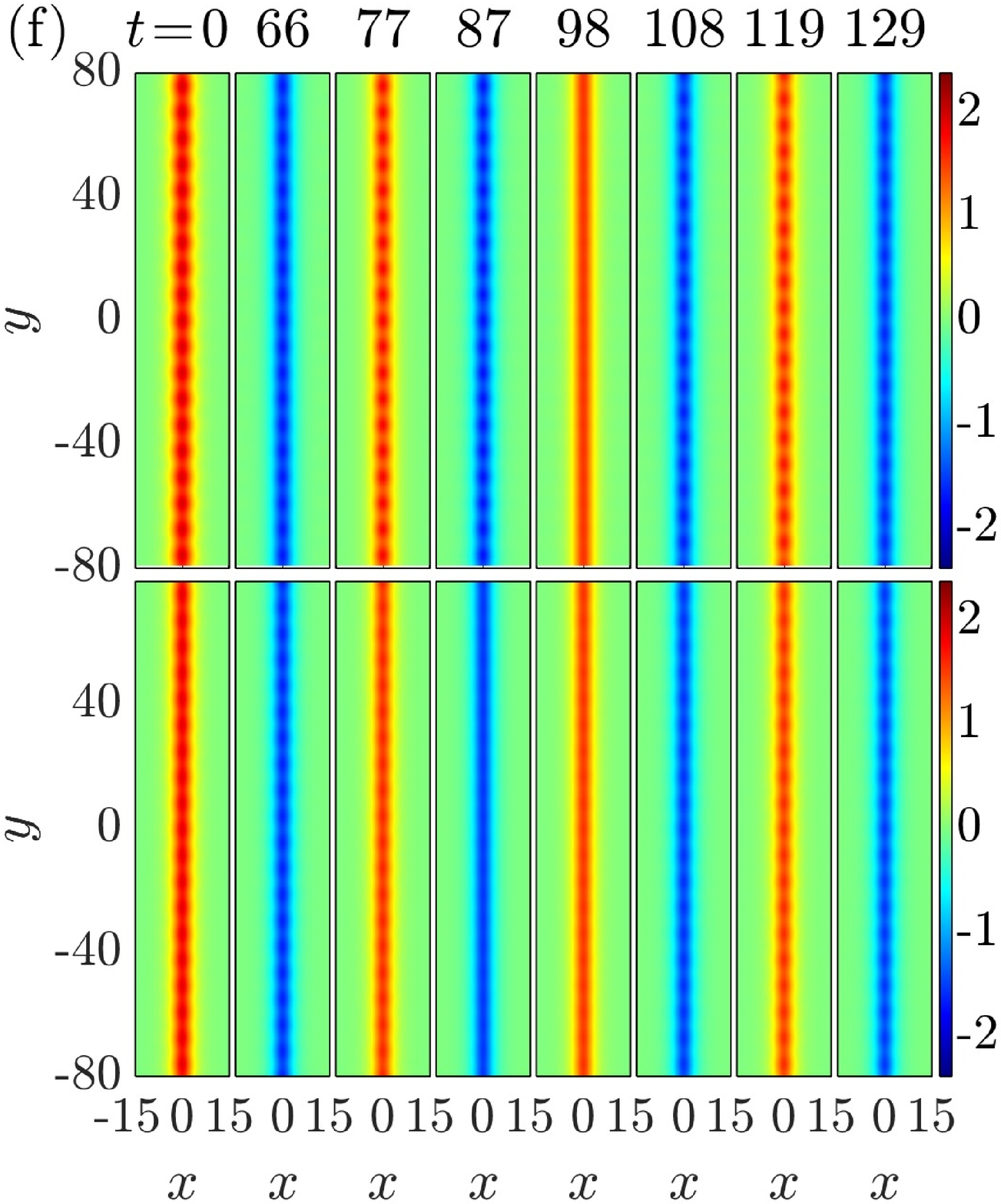}
\includegraphics[width=4.0cm]{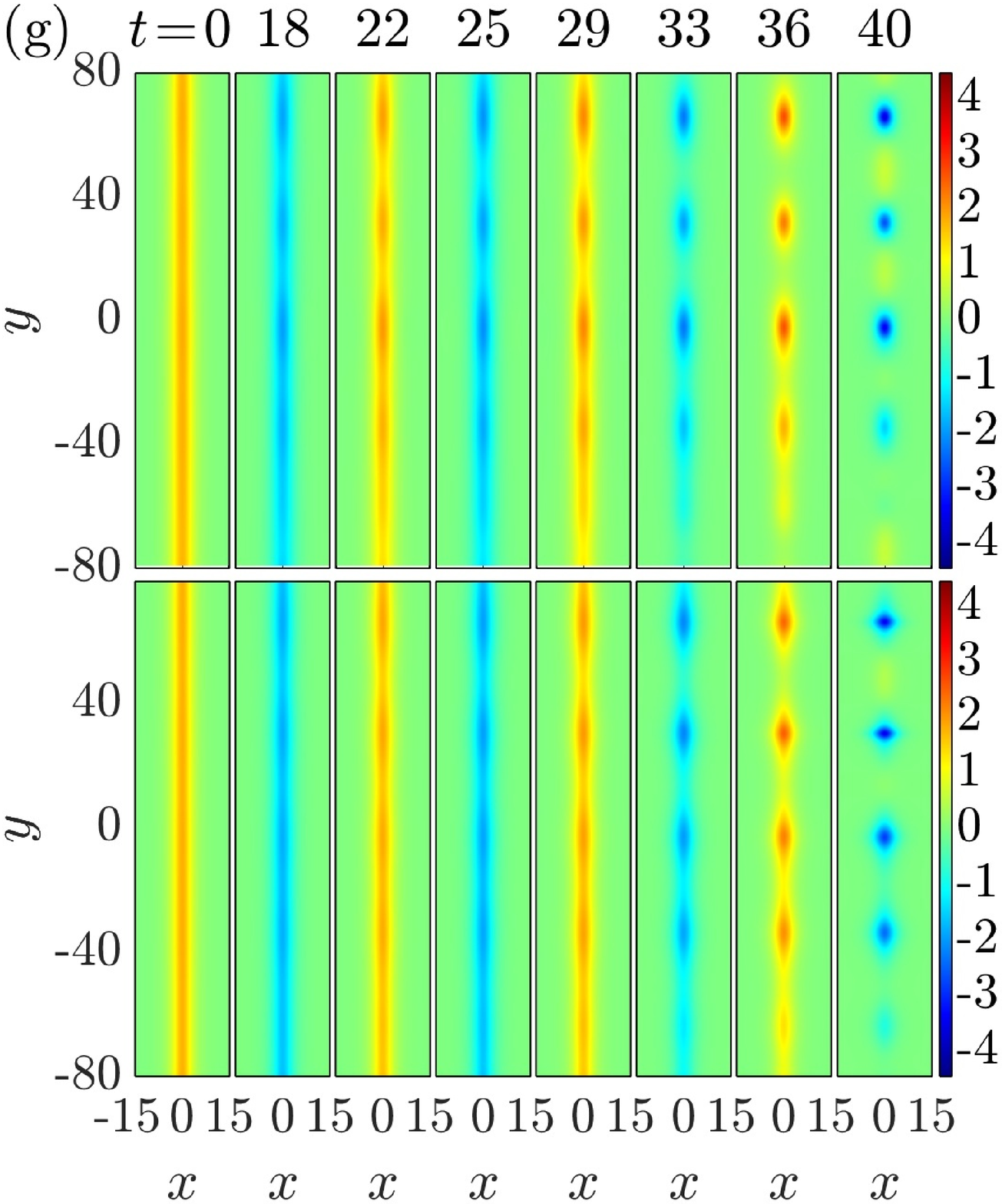}
\includegraphics[width=4.0cm]{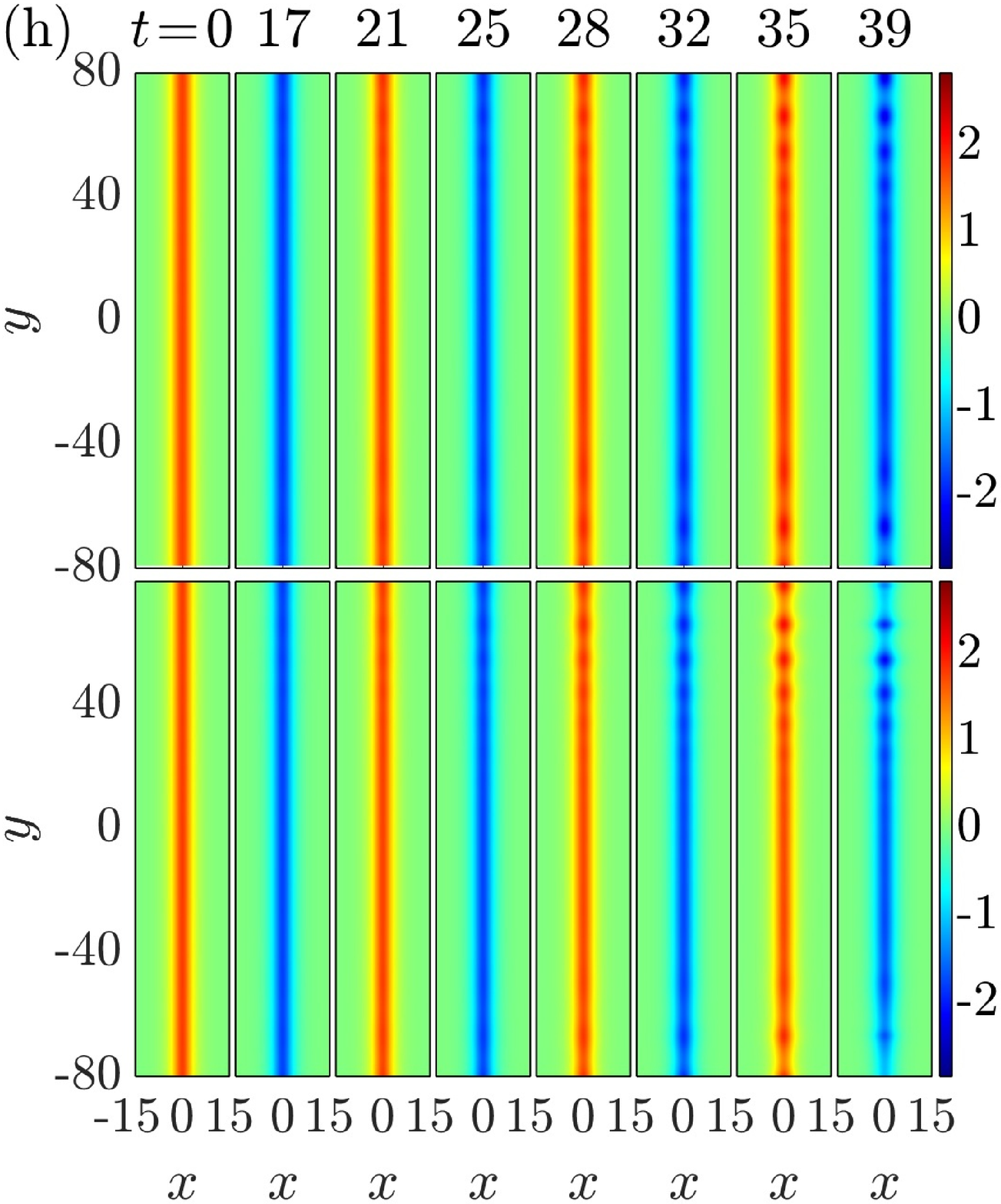}
\end{center}
\caption{(Color online)
Dynamics ensuing from the destabilization of the unstable breather stripe
under different perturbations.
Different panels, depicting $u(x,y,t)$, compare the development of
the modulational (necking) instability for the sG breather stripe, as
produced by the full sG model (in top subpanels), and by the VA
model given in Eq.~(\ref{eq:chi}) (in bottom subpanels)
for $\omega_{\mathrm{b}}=0.9$.
The input was a slightly perturbed exact sG
breather stripe, to which one [(a)--(f)] or several [(g) and (h)]
perturbation modes with wavenumbers $k_{y}$ are added.
The system was integrated in the domain $|x|\leq 40,|y|\leq 80$ (only the segment $|x|\leq 15$
is shown), with periodic boundary conditions along $y$ and homogeneous Dirichlet
boundary conditions along $x$.
Please see the main text for more details.
}
\label{fig:VA_dyn}
\end{figure*}

Before we compare the reduced dynamics, presented by Eq.~(\ref{eq:chi}), to
the evolution of the full sG breather stripe, we provide a generalization of
the VA methodology to incorporate the full nonlinearity of the system. If
instead of approximating $\sin (u)$ up to third order as done above, one
rewrites the nonlinearity in the effective Lagrangian as
$\sin^{2}(u/2)=(1/2)\left[ 1-\sum_{m=0}^{\infty }(-1)^{m}u^{2m}/(2m)!\right] $
and uses the identity,
\begin{equation}
\int_{-\infty }^{+\infty}\sech^{2m}(x)\,dx=2^{2m-1}\frac{\left[ \left(
m-1\right) !\right]^{2}}{\left( 2m-1\right) !},
\label{I}
\end{equation}
it is possible to evaluate the effective Lagrangian for the full sinusoidal
nonlinearity with ansatz (\ref{ans}). The respective extended Euler-Lagrange
equation is
\begin{equation}
\chi_{tt}-\chi_{yy}-\frac{\mu^{2}}{3}\chi -\sum_{m=1}^{\infty}
\frac{(-1)^{m}}{2}\left[ \frac{(m-1)!}{(2m-1)!}\right]^{2}(2\chi)^{2m-1}=0.
\label{eq:chi_sum}
\end{equation}
Although Eq.~(\ref{eq:chi_sum}) generalizes Eq.~(\ref{eq:chi}) in that it takes the
full nonlinearity into account, it does not provide an exact theory, because
the underlying ansatz~(\ref{ans}) is only valid for small amplitudes. Thus,
the extended equation may provide for a more accurate VA model of the
necking dynamics, but this model retains the
approximate nature of the above analysis (based on the proposed ansatz).

\begin{figure*}[htb]
\begin{center}
\includegraphics[height=5cm]{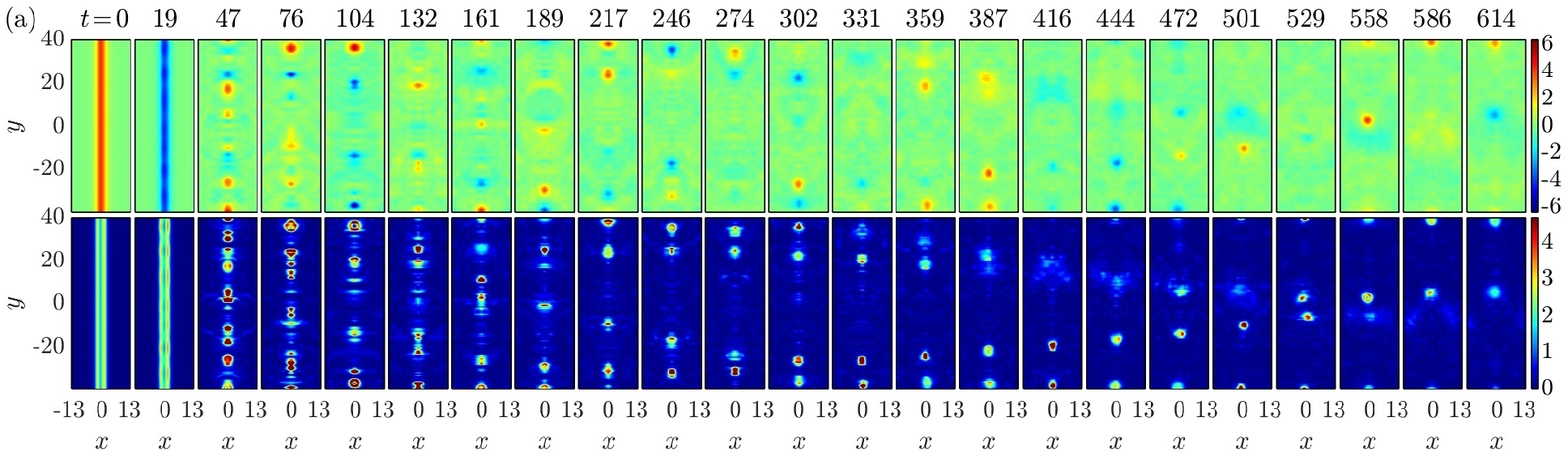}
\\[2.0ex]
\includegraphics[height=5cm]{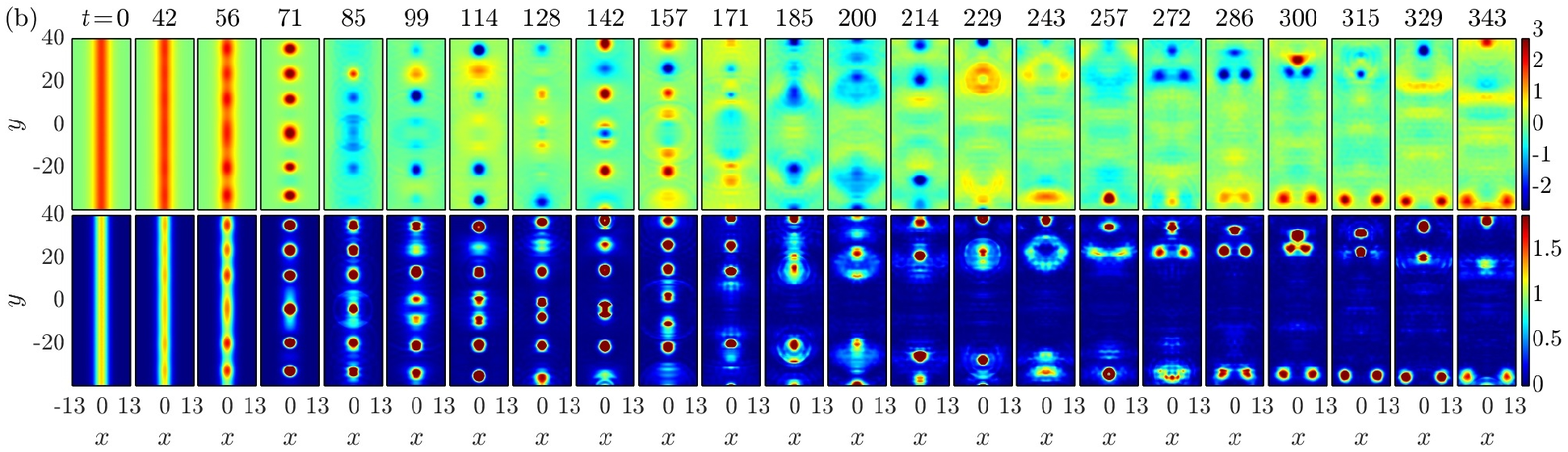}
\end{center}
\vspace{-0.3cm}
\caption{(Color online) Long term evolution of randomly perturbed breather stripes showing
the generation of ``blobs" (perturbed radial sG breathers) as a result of necking MI.
Depicted is the corresponding field $u(x,y,t)$ and energy density $\mathcal{E}(x,y,t)$
(top and bottom plots, respectively) at times indicated in the
panels for (a) $\omega_{\mathrm{b}}=0.5$ and (b) $\omega_{\mathrm{b}}=0.9$.
The input is
the exact sG breather stripe perturbed by a uniformly distributed random
perturbation of amplitude $0.01$. The domain size corresponds to
$|x|,|y|\leq 40$ (only the segment $-13\leq x\leq 13$ is shown).
}
\label{fig:sG_dyn}
\end{figure*}

To gain insight into the validity of the VA models~(\ref{eq:chi}) and
(\ref{eq:chi_sum}), in Fig.~\ref{fig:sG_vs_VA} we compare the stability spectra
provided by these approximations
with the numerically found main instability window of the
full sG model. Different panels correspond to replacing the infinite sum in
Eq.~(\ref{eq:chi_sum}) by the truncated sum $\sum_{m=1}^{N_{t}}$, that keeps
the first $N_{t}$ terms in the expansion. In this notation, Eq.~(\ref{eq:chi_sum})
reduces to Eq.~(\ref{eq:chi}) when $N_{t}=2$. The results
presented in Fig.~\ref{fig:sG_vs_VA} suggest that the MI is indeed adequately
captured by the VA in the region~(\ref{eq:01}).  This is true
qualitatively in that region and even quantitatively as
$\omega_{\mathrm{b}} \rightarrow 1$.
It is seen too that, as $1-\omega_{\mathrm{b}}$
increases, the match between the full sG model and its VA
counterparts deteriorates. Nonetheless, the VA reproduces the correct overall
trend and the shape of the instability spectra. As regards  keeping more terms in
the expansion of the nonlinearity in the Lagrangian, Fig.~\ref{fig:sG_vs_VA}
shows that proceeding from $N_{t}=2$ to $N_{t}=3$ does not produce a significant
improvement in the accuracy, and there is no discernible difference between
$N_{t}=4$ and $N_{t}>4$ either.
%
%
Therefore, from now on, we use the simplest reduced model~(\ref{eq:chi})
with $N_{t}=2$ for the comparison with the full sG model.

\subsection{Dynamical evolution of breather stripes}

Having tested the reduced models in terms of the stability spectra, we now
extend the comparison by numerically following the necking dynamics in the
framework of the full sG model and the VA for a number of scenarios. The
evolution was initiated by inputs
\begin{eqnarray}
u(x,y,0)&=&u_{\mathrm{b}}(x,0)\,
\left[1+\varepsilon \cos\left(\frac{n_{y}\pi}{L_{y}}y\right)\right],
\notag
\\
u_{t}(x,y,0)&=&0,
\label{eq:ny}
\end{eqnarray}
where $n_{y}$ is the perturbation's integer mode number (associated
with the wavenumber $n_y \pi/L_y$) and $\varepsilon$
its relative strength, while $u_{\mathrm{b}}(x,0)$ is the exact 1D sG
breather profile~(\ref{eq:1D_breather}). Results of the comparison are
depicted in Fig.~\ref{fig:VA_dyn}.
The dynamics is depicted by means of surface plots
of $u(x,y,t)$ at the indicated times.
The input was a slightly perturbed exact sG
breather stripe, to which one [(a)--(f)] or several [(g) and (h)]
perturbation modes with wavenumbers $k_{y}$ were added as follows.
The perturbations were constructed by seeding the necking modulation as per
Eq.~(\ref{eq:ny}), with perturbation strength $\varepsilon =0.01$. Panels
(a)--(d) depict the comparison for the perturbations with small wavenumbers
$n_{y}=1$, $3$, $5$, and $7$, corresponding to the necking MI. Panels (e) and
(f) correspond to large perturbation wavenumbers, with $n_{y}=17$ and
$n_{y}=19$, which are taken, respectively, just below and above the
instability threshold. For the stable case (f) with $n_{y}=19$, a stronger
perturbation with $\varepsilon =0.1$ was used to stress the
stability of the sG breather against this perturbation wavenumber. Finally,
panels (g) and (h) correspond to cases when a mix of perturbation modes with
random strengths (chosen in interval $-0.01\leq \varepsilon \leq 0.01
$) is introduced in the initial conditions. Case (g) contains perturbations
with modes $1\leq n_{y}\leq 5$, while in case (h) we used $1\leq n_{y}\leq 22$.
In all the cases, to match the underlying oscillation frequencies of the
exact breather stripe and its VA counterpart, the time for the VA results is
scaled by a fitting factor. This factor, for all the cases shown here,
amounts to a reduction of the time between $8$ and $10\%$.
%
%
As seen in the figure, and as predicted by the stability spectra, perturbations with
small wavenumbers always lead to necking MI. The perturbation eventually breaks the
stripe into a chain of self-trapped ``blobs" (localized modes). These
time-oscillating ``blobs" are identified as highly perturbed radial breathers, this
conclusion being the second key aspect of our study (see the following
Section~\ref{sec:radial}, and also Ref.~\cite{caputo}). All the modes with
perturbation indices $1\leq n_{y}\leq 17$ ($k_{y}=n_{y}\pi /L_{y}$) [see
Eq.~(\ref{eq:ny})] are unstable, and the VA is able to accurately capture the
corresponding instability and ensuing dynamics, see panels (a)--(e). Naturally, as the
necking perturbation grows, the VA becomes less accurate, as the underlying
variational ansatz~(\ref{ans}) is no longer an appropriate approximation for
the profile of the deformed stripe. Nonetheless, for times up to those at
which the stripe breaks into a chain of ``blobs", the VA
evolves in tandem with the full sG dynamics. This is also true when, instead
of using a single mode as the perturbation, one perturbs the stripe by a
collection of modes, as shown in panels (g) and (h) where the perturbations
contain, respectively, a combination of the lowest $5$ modes and even as
many as $22$ ones. Finally, in panel (f) of Fig.~\ref{fig:VA_dyn} we present
an example of the evolution with the perturbation index ($n_{y}=19$) chosen
beyond the instability boundary of the main MI window. In this stability
case, the simulations demonstrate that the full sG dynamics is closely
followed by the VA for longer times, in comparison to the unstable cases.

\begin{figure}[htb]
\begin{center}
\includegraphics[width=3.0cm]{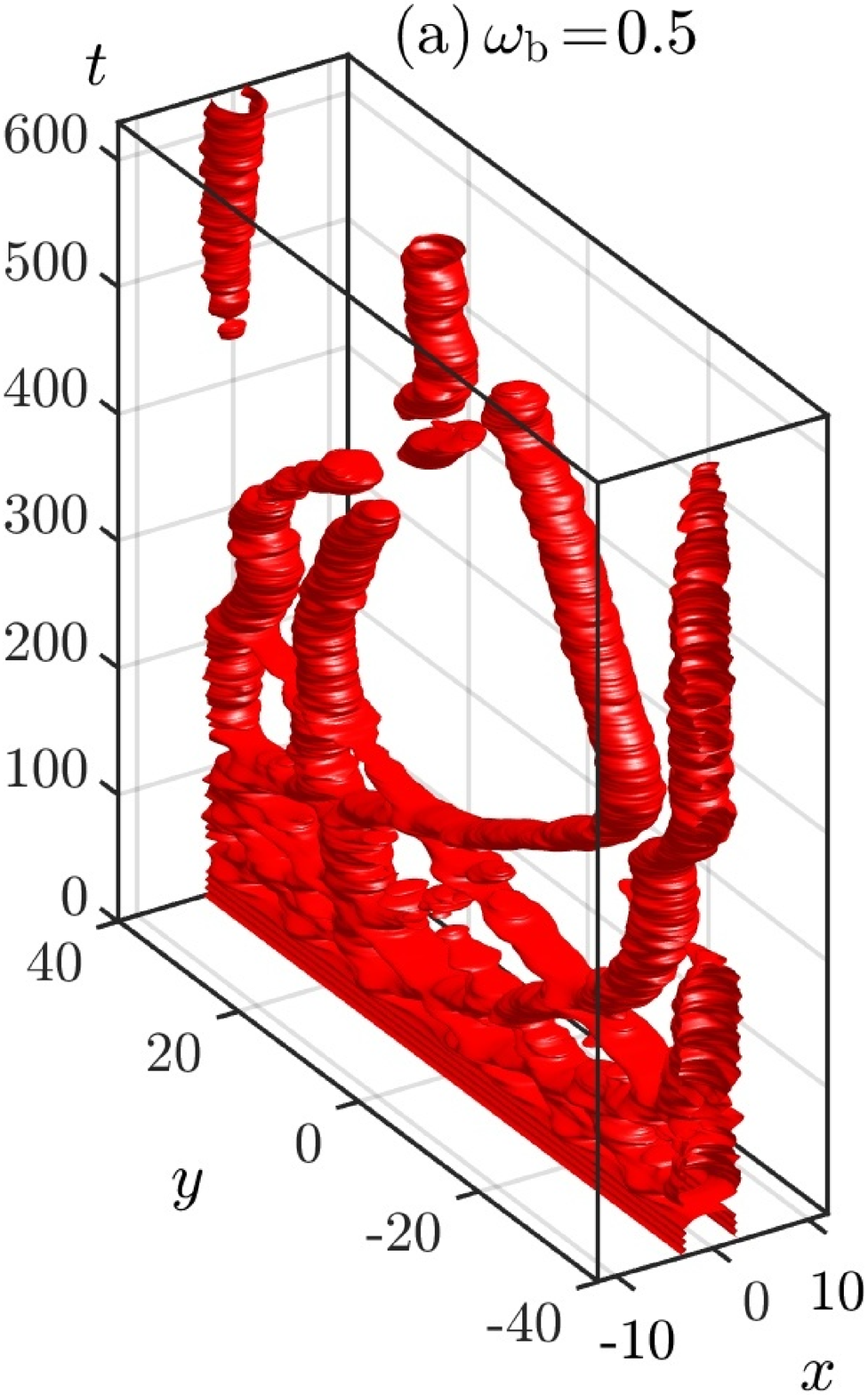}
\quad
\includegraphics[width=3.0cm]{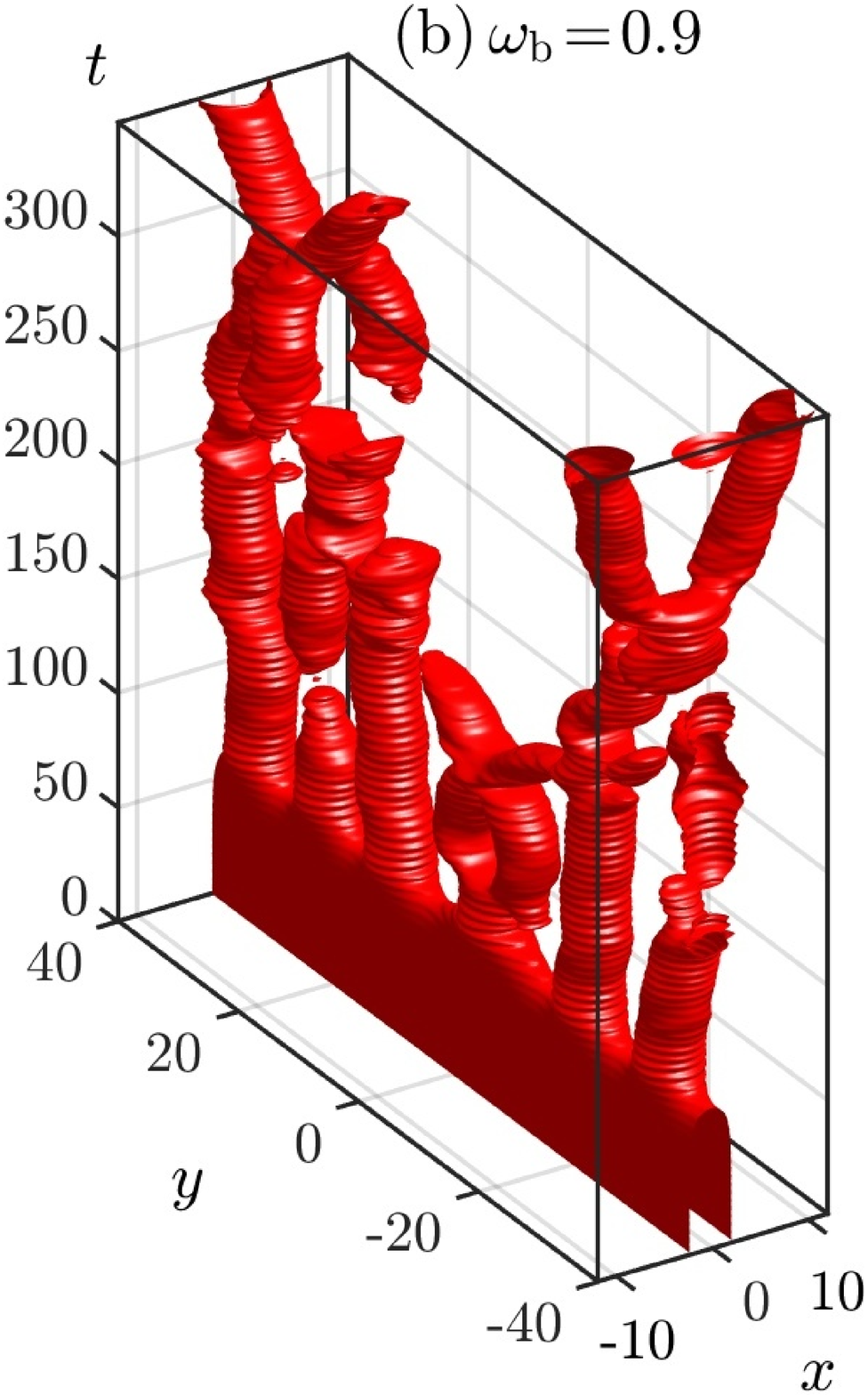}
\end{center}
\vspace{-0.3cm}
\caption{(Color online) Isocontour plots of the energy density $\mathcal{E}(x,y,t)$
corresponding to the evolution of slightly perturbed sG breather
stripes shown in Fig.~\ref{fig:sG_dyn}. In both cases, random
perturbations added to the initial stripe trigger the development of its
transverse MI. As a result, several robust ``blobs" emerge,
whose motion and interactions dominate subsequent dynamics.}
\label{fig:sG_dyn_3D}
\end{figure}

Figure~\ref{fig:sG_dyn} expands on the results presented in Fig.~\ref{fig:VA_dyn}
by considering random perturbation of the breather stripes and allowing the dynamics
to evolve for longer times. We observe that, similarly to Fig.~\ref{fig:VA_dyn},
Fig.~\ref{fig:sG_dyn} shows the formation
of long-lived spatially localized ``blobs". The top row of panels depicts
snapshots of the field
$u(x,y,t)$ at the indicated moments of time. Since the solutions that we are
inspecting are oscillatory in time, in the corresponding bottom row we plot
snapshots of the corresponding 2D energy density,
\begin{equation}
\mathcal{E}(x,y,t)=\frac{1}{2}\left( u_{t}^{2}+u_{x}^{2}+u_{y}^{2}\right) +V(u),
\label{eq:energy}
\end{equation}
which allows us to partially eliminate the oscillations, and thus to better
capture the emergence and dynamics of the ``blobs"
(localized breathers). The figure suggests that the breathers generated by
the MI from the unstable stripes are generically robust, although some of
them disappear, while others spontaneously emerge from condensation of the
background random field. Note that the localized breathers travel and, as a
result, collide or merge with neighboring ones. In Fig.~\ref{fig:sG_dyn_3D},
we display 3D isocontour plots of the energy density in the $(x,y,t)$
spatiotemporal continuum. This figure makes it possible to better follow the
emergence, evolution, and interactions of the ``blobs".

\section{Radial Breathers}
\label{sec:radial}

We now turn to the consideration of breathers with axial symmetry, i.e.,
$u(x,y,t)=u(r,t)$, with $r=\sqrt{x^{2}+y^{2}}$ in a circular domain of radius
$R$. Their existence is suggested by the findings reported in the previous
Section, where necking MI splits the sG breather stripe into a chain of persistent
localized modes (``blobs").

\begin{figure}[htb]
\begin{center}
\includegraphics[height=4.5cm]{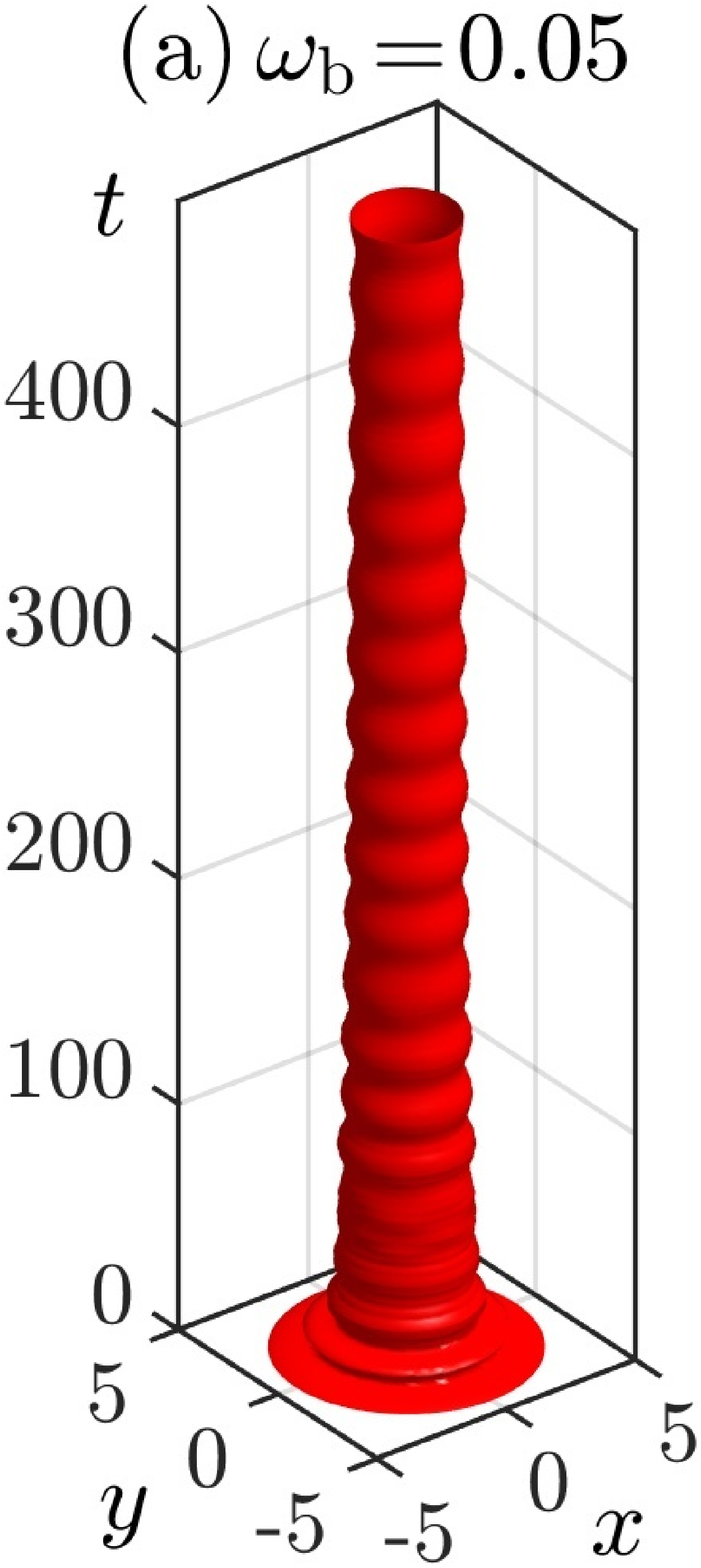}
\includegraphics[height=4.5cm]{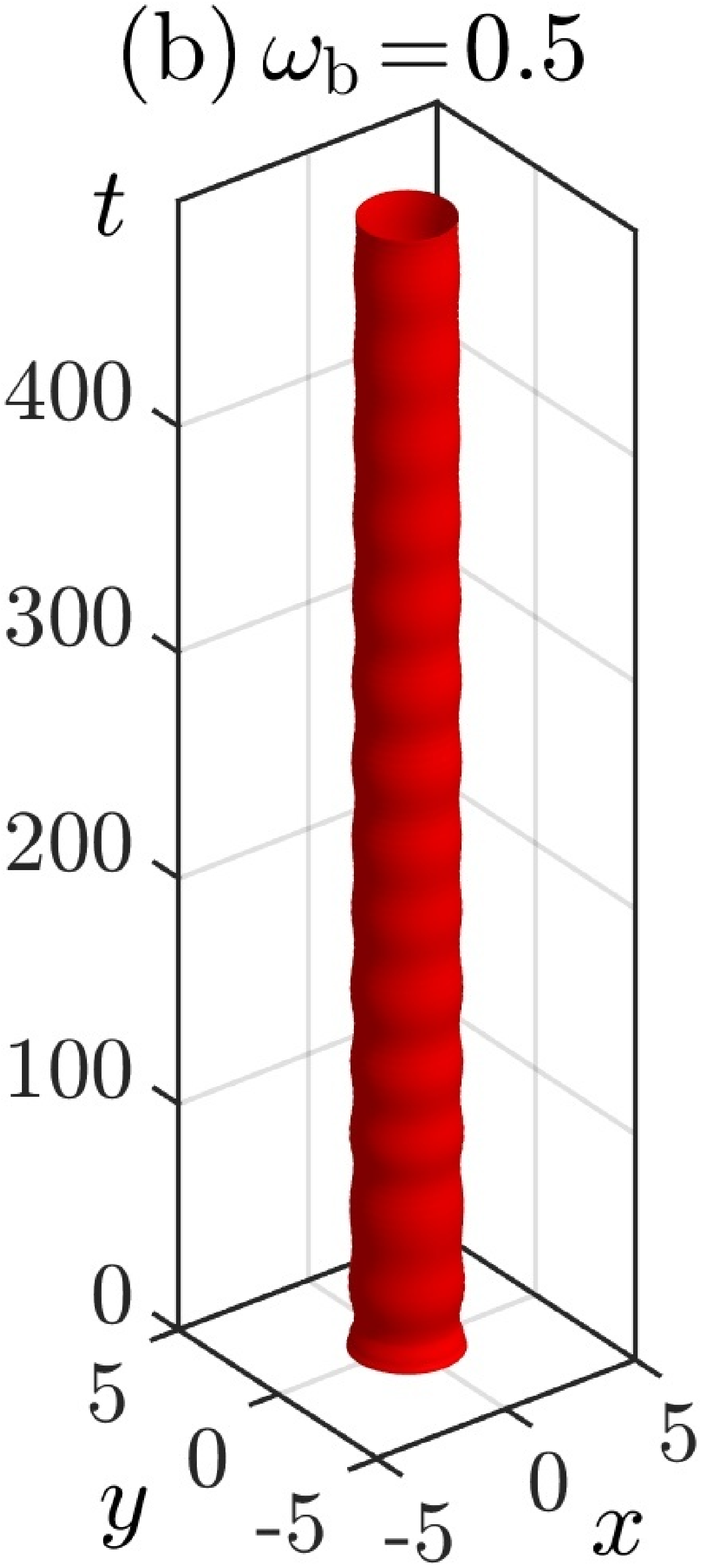}
\includegraphics[height=4.5cm]{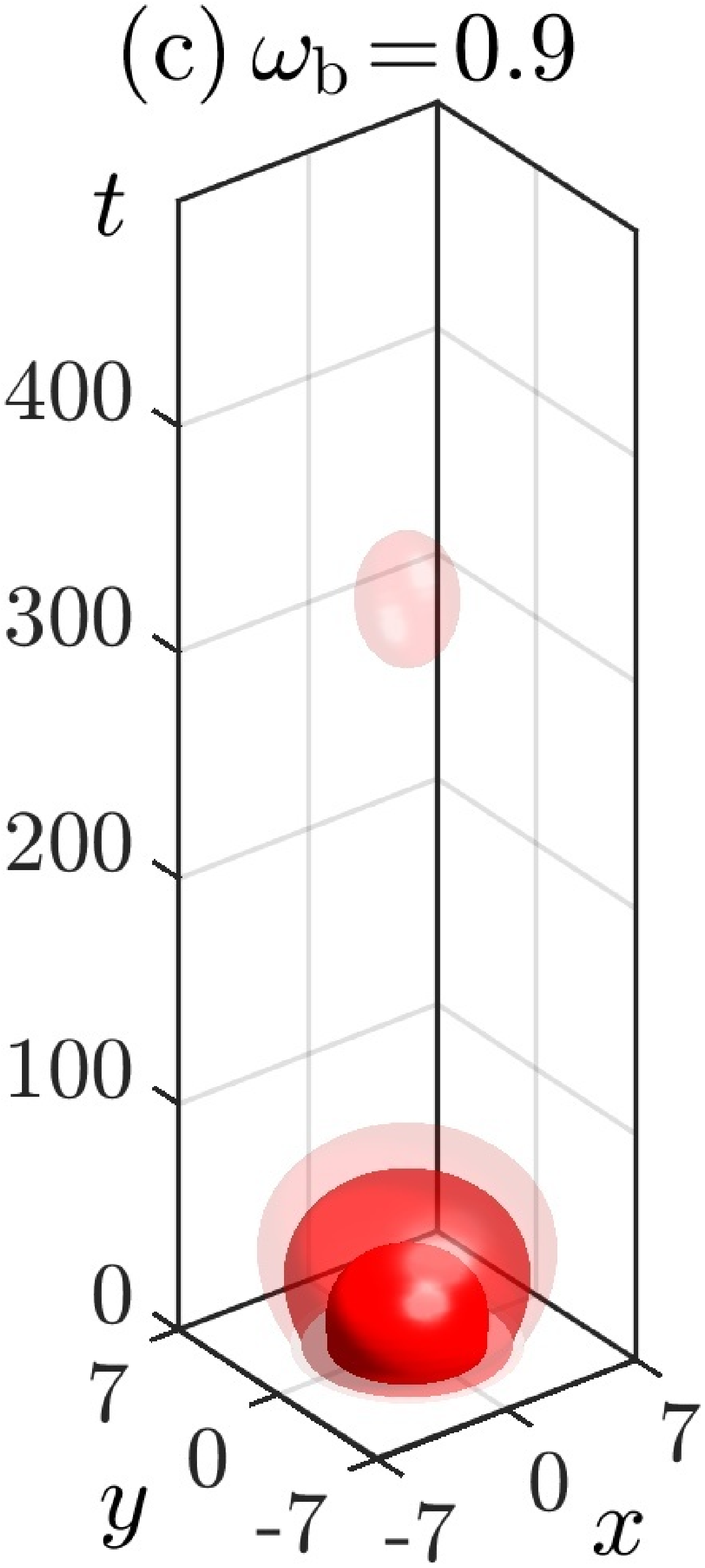}
\includegraphics[height=4.5cm]{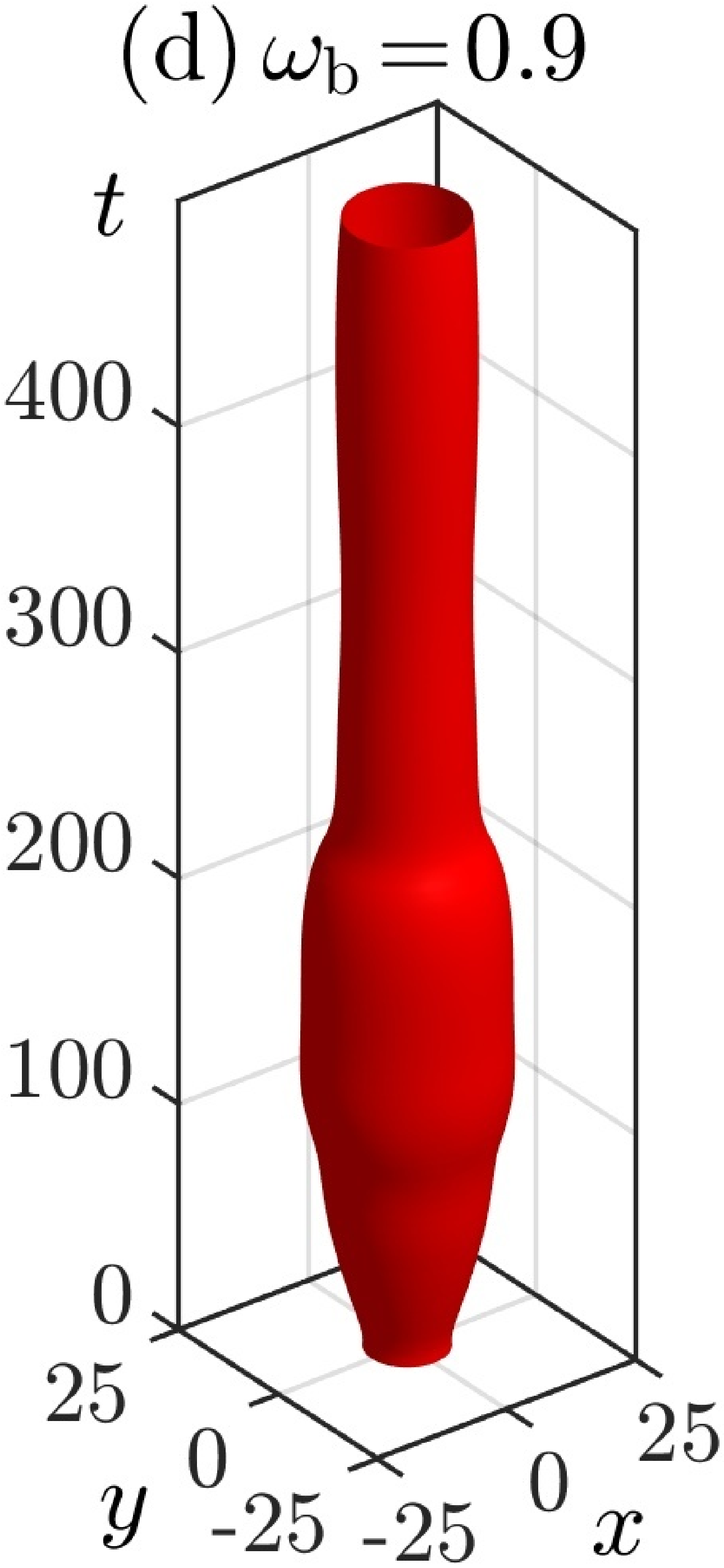}
\end{center}
\caption{(Color online) Isocontour plots depicting the evolution of the energy
density $\mathcal{E}(x,y,t)$ for radial waveforms initialized by the 1D
breather profile, with $x$ replaced by $r$, for (a) $\omega_{\mathrm{b}}=0.05$,
(b) $\omega_{\mathrm{b}}=0.5$, and (c)-(d) $\omega_{\mathrm{b}}=0.9$.
The waveforms reshape themselves into
radial breathers, which persist over an indefinitely large number of
oscillations. In (a) and (b), the isocontour levels correspond to
$\mathcal{E}(x,y,t)=\mathcal{E}_{0}$, with $\mathcal{E}_{0}=2.5$; in (c), three cuts are shown for
$\mathcal{E}_{0}=0.1$, $0.15$, and $0.3$ (more transparent isocontours correspond to
smaller $\mathcal{E}_{0}$). Apparently, the radial breather for $\omega_{\mathrm{b}}=0.9$
seems to disperse. However, an isocontour drawn for an even
smaller value, $\mathcal{E}_{0}=0.02$, in (d) suggests the presence of a much
shallower radial breather.
}
\label{fig:sG_dyn_radial_3D}
\end{figure}

\begin{figure}[htb]
\begin{center}
\includegraphics[height=4.5cm]{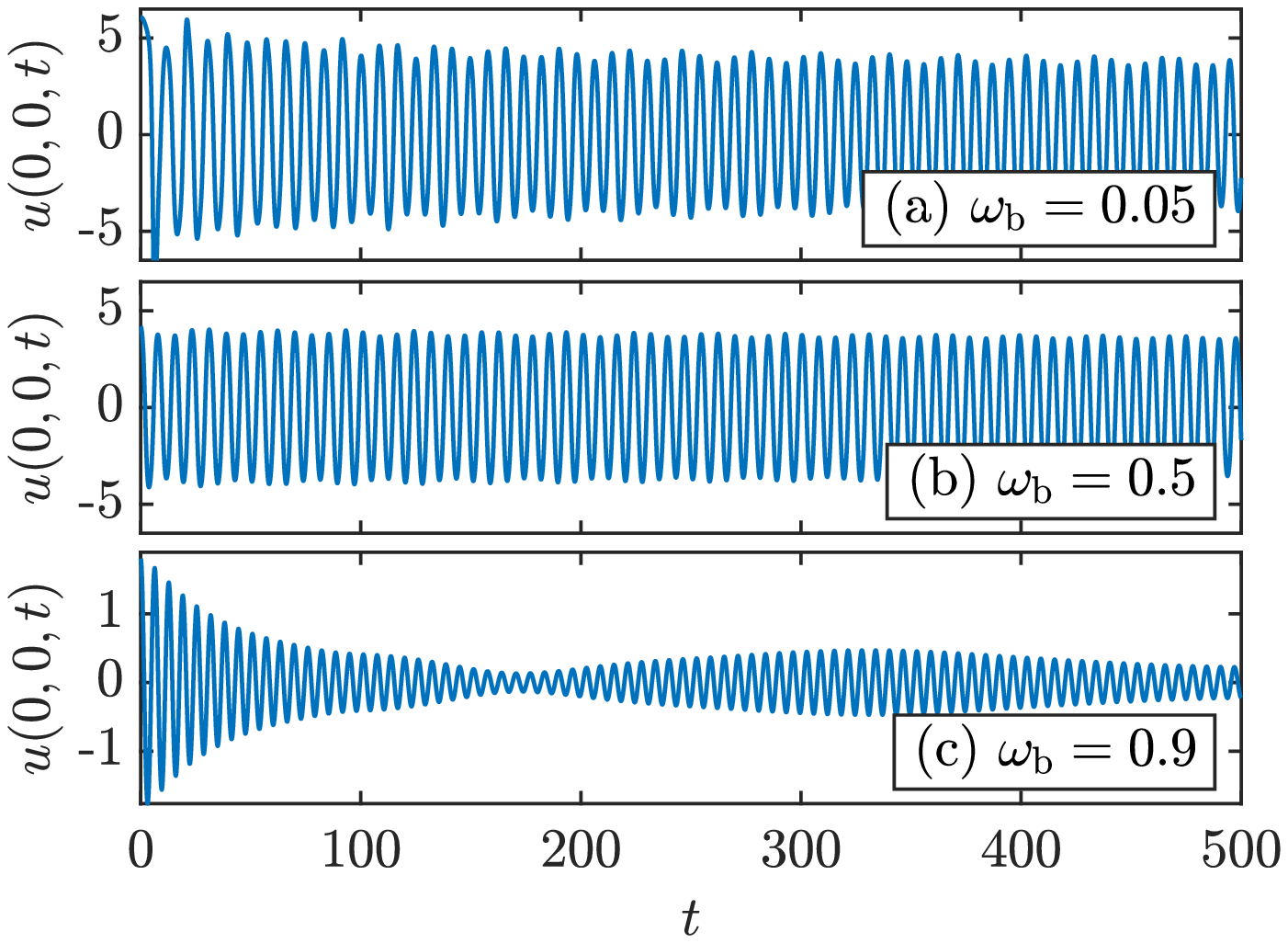}
~~~
\includegraphics[height=4.5cm]{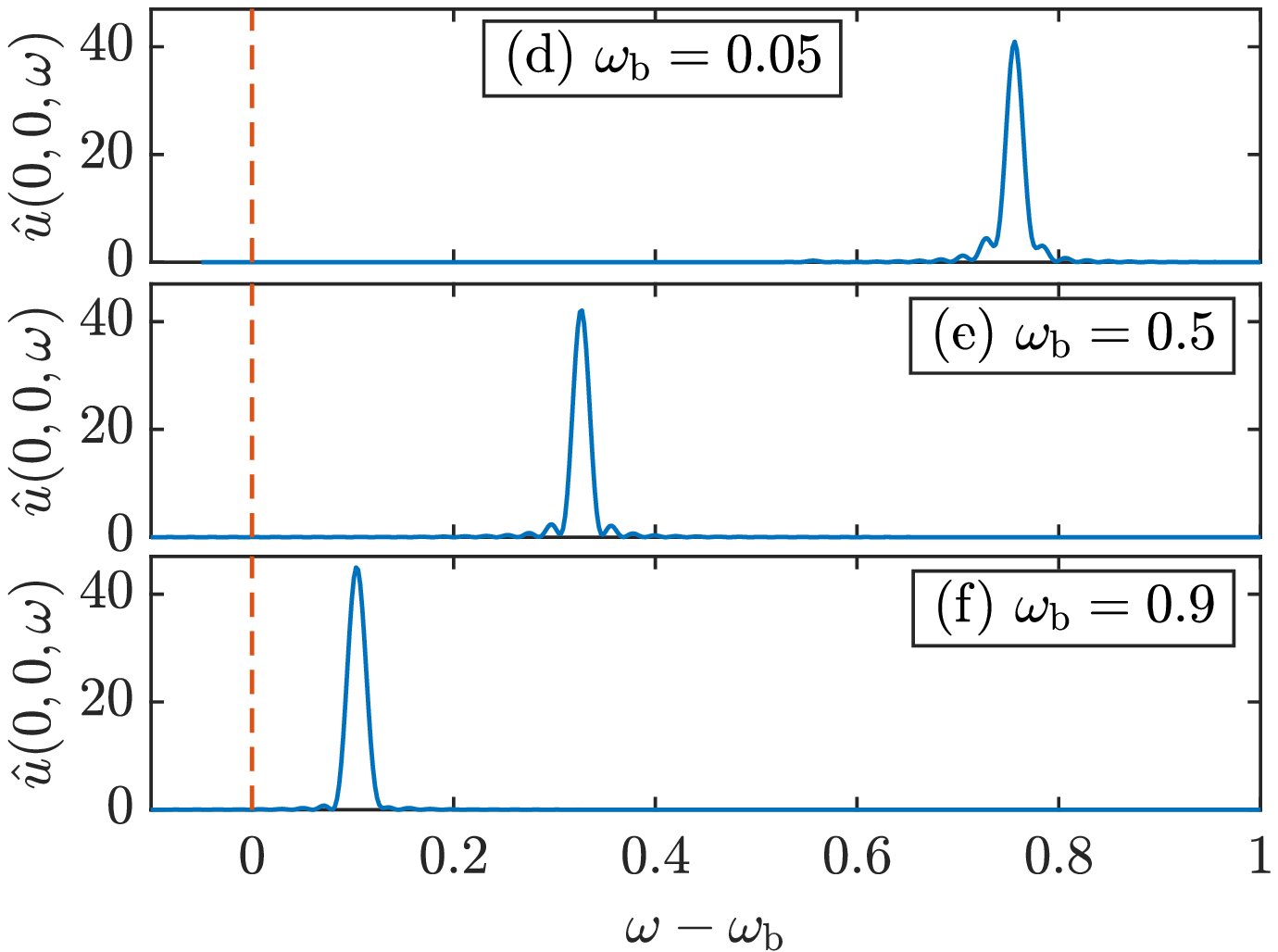}
\end{center}
\vspace{-0.4cm}
\caption{(Color online) Time series depicting the evolution of the field at the
central point $u(0,0,t)$ (top panels) and the corresponding normalized power
spectrum density (PSD) $\hat{u}(0,0,t)$ for $200\leq t \leq 500$
(bottom panels), for the cases shown in Fig.~\ref{fig:sG_dyn_radial_3D}.
For the cases with
$\omega_{\mathrm{b}}=0.05$ (a) and $\omega_{\mathrm{b}}=0.5$ (b) the
resulting radial breathers are very similar in terms of the field amplitude
[see plots for $u(0,0,t)$] and dominant frequency [see plots for $\hat{u}(0,0,\omega )$].
In panel (c), corresponding to the case with
$\omega_{\mathrm{b}}=0.9$, a much shallower radial breather emerges
with a dominant frequency tending to $\omega =1$, which is the edge
of the phonon band. The corresponding dominant frequencies are
$\omega \approx 0.8027$ in (a), $\omega \approx 0.8215$ in (b), and
$\omega \approx 0.9986$ in (c).
Red vertical dashed lines denote
frequencies $\omega_{\mathrm{b}}$ of the 1D sG breather used as the
input. }
\label{fig:sG_dyn_radial_series}
\end{figure}

\begin{figure}[htb]
\begin{center}
\includegraphics[width=7.0cm]{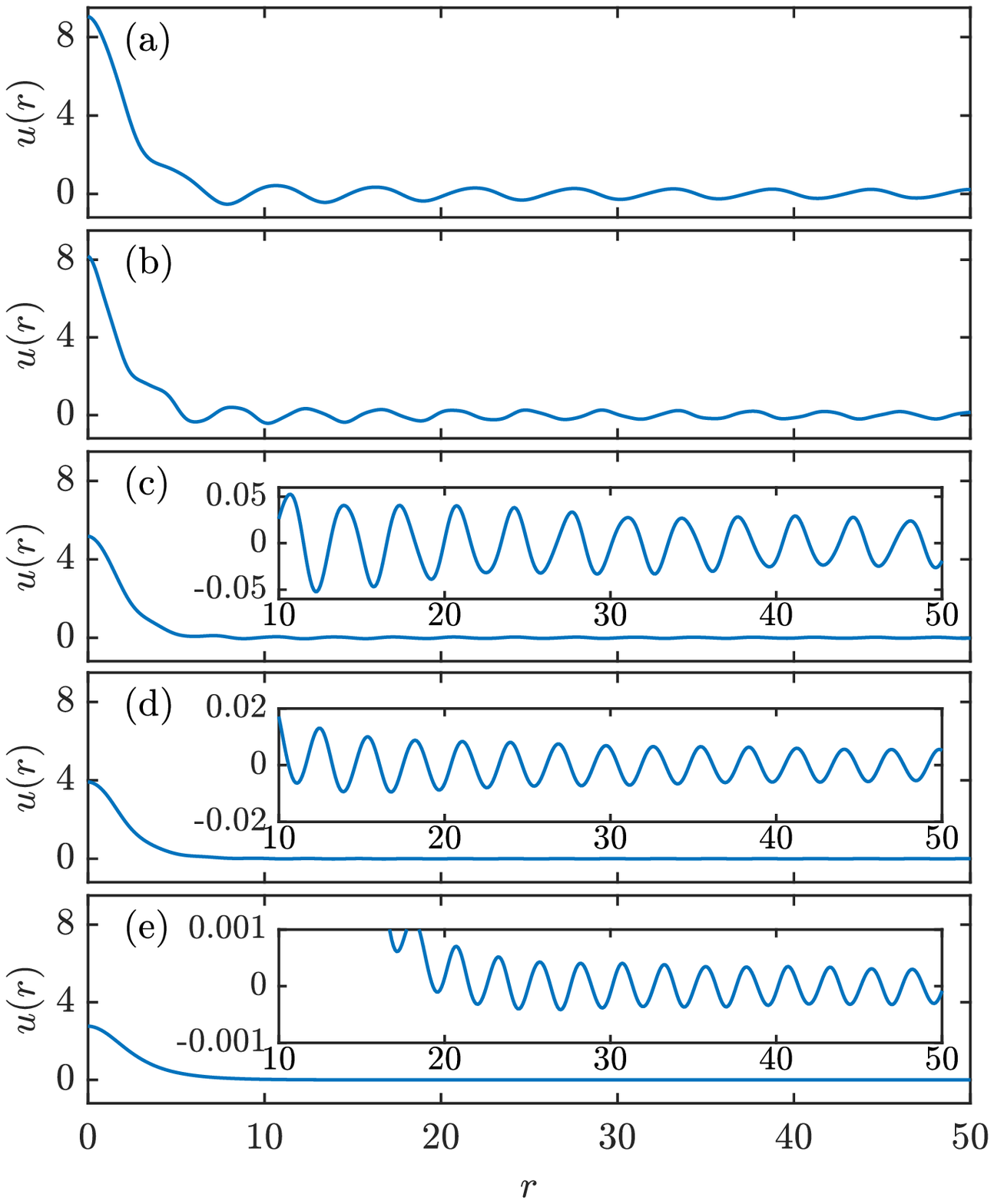}
\end{center}
\vspace{-0.4cm}
\caption{(Color online) Numerically computed profiles of radial breathers for
$\omega_{\mathrm{b}}=0.5,0.6,0.7,0.8,0.9$ (from top to bottom). The
solution domain is $0\leq r\leq 200$, only the central portion $r\leq 50$
being shown. All these profiles have oscillatory tails with a small
amplitude (see insets for $\omega_{\mathrm{b}}\in {0.7,0.8,0.9}$). }
\label{fig:radial_breathers}
\end{figure}

\subsection{Constructing radial breathers}

With the aim to construct initial conditions that directly lead to radial
breathers, we take, as an input, the exact 1D breather~(\ref{eq:1D_breather}),
with the $x$ coordinate replaced by the radial one:
$u_{\mathrm{radial}}(r,0)=u_{\mathrm{b}}(r,0)$. The results are summarized in
Figs.~\ref{fig:sG_dyn_radial_3D} and \ref{fig:sG_dyn_radial_series}, which depict,
respectively, isocontours of the energy density, and dynamics and power
spectra of the field at the central point, $u(0,0,t)$. The results suggest
that persistent radial breathers are generated by appropriately crafted
inputs. However, the resulting radial breathers do not keep the oscillation
frequency corresponding to the one seeded by the initial condition. In
fact, for $\omega_{\mathrm{b}}=0.05$ and $\omega_{\mathrm{b}}=0.5$, the
ensuing radial breathers are very similar (in amplitude and frequency), but
they feature an oscillation frequency higher than the one introduced by the
input, see green vertical dashed lines in Figs.~\ref{fig:sG_dyn_radial_series}(d)--(f).
Note that, as observed in panel (f) of Fig.~\ref{fig:sG_dyn_radial_series},
the input with larger $\omega_{\mathrm{b}}$
leads to a much shallower radial breather with a dominant frequency close to
$1$, i.e., the edge of the phonon band. Similar results were obtained for
$\omega_{\mathrm{b}}=0.75$ and $\omega_{\mathrm{b}}=0.8$ (not shown here).

The numerical experiments shown above might suggest that radial breathers
exist as exact solutions. However, on the contrary to the 1D sG equation,
its 2D counterpart is not integrable. Therefore, genuine radial breathers
cannot exist in the infinite spatial domain, as multiple breathing
frequencies, $n\omega $ with $\omega <1$ but $n>1/\omega $, resonate with
the frequencies belonging to the phonon spectrum, $\omega >1$, which should
give rise to the radiative decay of the breather (only integrable equations make
an exception, nullifying the decay rate)~\cite{eilbeck,dauxois,kivsharmalomed}.
However, in some cases, accurate simulations show that the rate
of the radiative decay is so small that the 2D sG breathers survive
thousands of oscillation periods, with a negligible loss of amplitude~\cite{geicke}.
Nevertheless, if the domain is \emph{finite}, there will be
gaps in the spectrum, which admit the existence of intraband breathers
similar to the \textit{phantom breathers} in discrete lattices introduced in
Ref.~\cite{phantom} and feature the aforementioned \textit{nanopteron} wings
oscillating in space. For example, we plot in Fig.~\ref{fig:radial_breathers}
profiles of numerically exact radial breathers with frequencies
$\omega_{\mathrm{b}}={0.5,0.6,0.7,0.8,0.9}$, computed in the radial domain $0\leq
r\leq 200$ and with zero initial velocity $u_t(r,t=0)=0$
(see below for more details of these numerics). As is made
evident by the figure, the radial breathers indeed have distinctive
oscillatory tails, which are more prominent for lower values of
$\omega_{\mathrm{b}}$, and their amplitude is decaying to zero, in line with the
expectation predicted above by the NLS reduction, in the limit of
$\omega_{\mathrm{b}}\rightarrow 1$. Indeed, the 2D NLS equation with the cubic
self-focusing term creates radial 2D \textit{Townes solitons}; see, e.g.,
Refs.~\cite{Fibich,bm} for a review and Refs.~\cite{moll,chen} for their
experimental realizations, an earlier one in optics and a recently created
Townes soliton in atomic gases. Although 1D breathers are also intraband
ones and possess a wing in the case of the finite-domain computation, the
continuation of 2D radial breathers within the band is a much more subtle
problem. A full analysis of the continuation (bifurcation) scenario in the
whole range of available frequencies, for which the non-integrability for
the 2D equation must play a crucially important role, is outside the
scope of the present work.
Therefore, in the  summary of the stability results that we present
below, we focus on branches of radial breathers for large
$\omega_{\mathrm{b}}$, \textit{viz}., $\omega_{\mathrm{b}}>0.899$, in
line with our earlier consideration of the limit of
$\omega_{\mathrm{b}} \rightarrow 1$. An
advantage of this option is also that, when dealing with larger
$\omega_{\mathrm{b}}$, the numerics (performed over one period through the Floquet
analysis, see below) are more manageable (faster) than for smaller frequencies.

\begin{figure}[tb]
\begin{center}
\includegraphics[width=8.0cm]{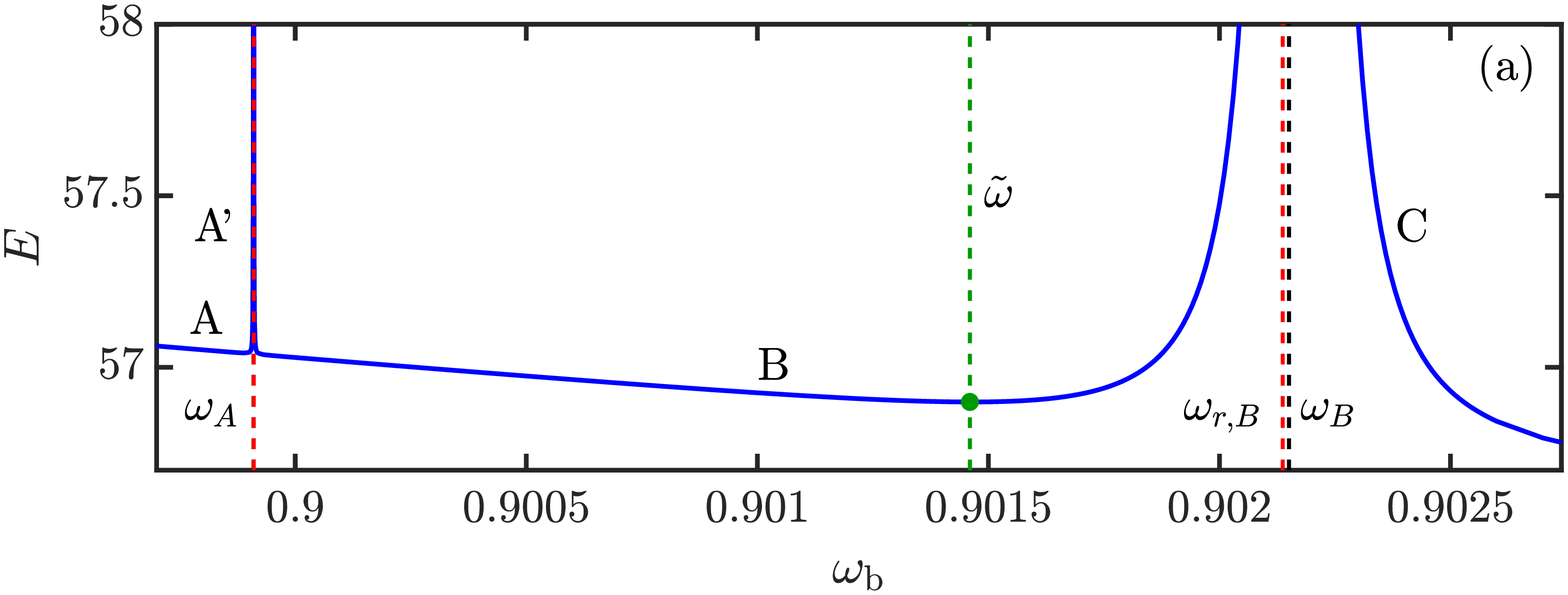}
\\[3.0ex]
\includegraphics[width=8.0cm]{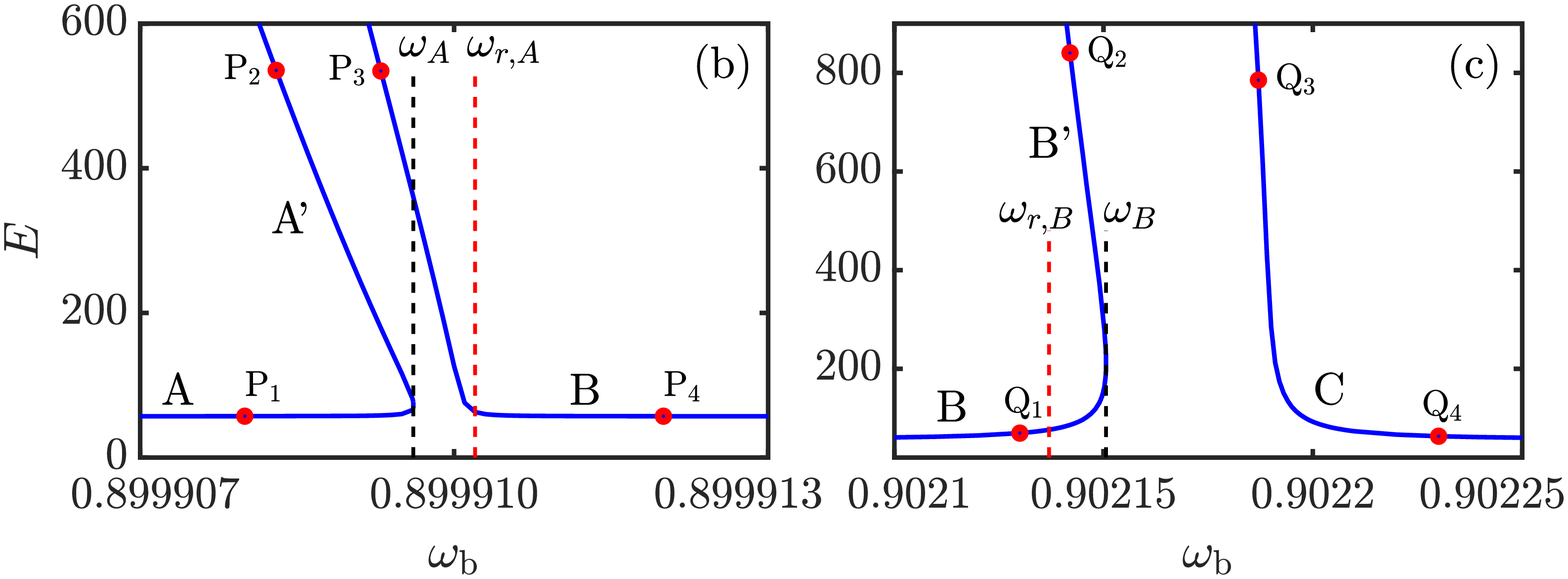}
\\[2.0ex]
\includegraphics[width=4.5cm]{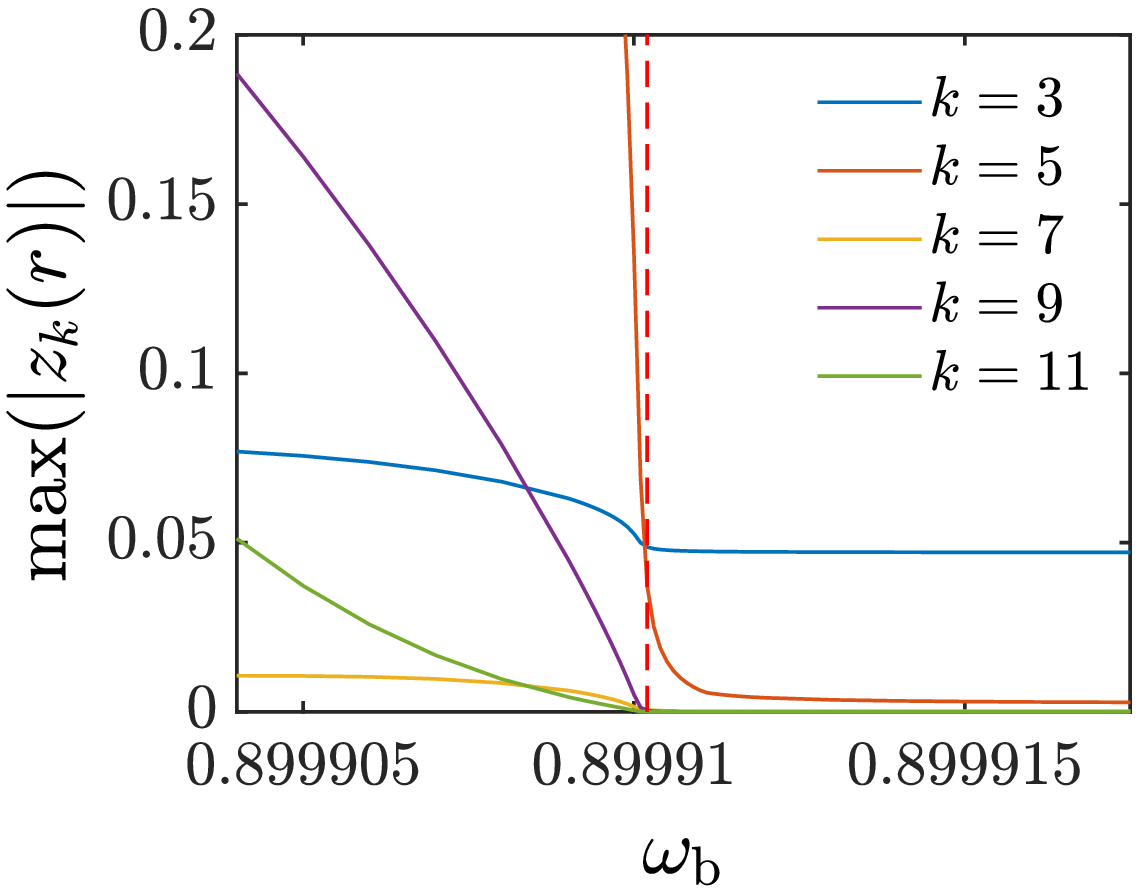}
\end{center}
\caption{(Color online) Existence branches of radial breathers for
$\omega_{\mathrm{b}}$ close to $1$. (a--c): Total energy $E$ of radial
breathers versus the breather frequency, $\omega_{\mathrm{b}}$.
Panel (a) presents an overview of the different solution branches (see the
text), while panels (b) and (c) display zoomed in versions of the transition zones
between A-B and B-C branches, respectively.
Representative profiles for the solutions
belonging to each branch, corresponding to the large red dots, are presented in
Figs.~\ref{fig:profilesAB} and \ref{fig:profilesBC}.
Vertical dashed lines denote the most relevant frequency values:
(i) dashed red lines, corresponding to $\omega_{r,A}$ and
$\omega_{r,B}$, denote the interval outside of which the breather acquires a
minimum at $r=0$;
(ii) dashed black lines, corresponding to $\omega_{A}$ and $\omega_{B}$,
denote the turning points for branches A and B, respectively;
(iii) the dashed green line at $\omega_{\mathrm{b}}=\tilde{\omega}$
denotes the location of the energy
minimum for branch B, above which the breathers belonging to this branch are
prone to exponential instabilities.
%
%
The bottom
panel depicts the maximum amplitude of the main Fourier coefficients of the
breathers belonging to branch B.
All the computations were performed in region $r\leq R=200$. }
\label{fig:breather_energy}
\end{figure}

\subsection{Radial breathers solution branches and stability}

We now consider the stability of a ``radial breather'' $u_b(r,t)$.
For this purpose, we proceed as in the breather stripes section. First,
we write the PDE (\ref{eq:sG}) in polar coordinates and obtain
the equation for the perturbation $w$ corresponding to Eq.~(\ref{wtt}):
\begin{equation}
w_{tt} - {1\over r} (r w_r)_r - {1\over r^2}w_{\theta\theta} + \cos(u_b(r,t)) w =0 .
 \label{wtta}
\end{equation}
We then assume a solution
\begin{equation}
w (r,\theta,t)=\zeta (r,t)\exp (i k_{\theta} \theta ),
\label{wkya}
\end{equation}
with integer angular wavenumber $k_{\theta}$,
yielding the final time-periodic PDE for the perturbation $\zeta$:
\begin{equation}
 \zeta_{tt} - {1\over r} (r \zeta_r)_r + \left ( {k_{\theta}^2 \over r^2} + \cos(u_b(r,t)) \right ) \zeta =0.
\label{zett}
\end{equation}
To deal with the formal singularity of the radial Laplacian at
$r=0$ we do not include this point in the domain
so that the closest point to $r=0$ is $r=h/2$ [that
is, $r\in (h/2,R)$], and assume $u_r(r=0) =0$.
This way of handling the singularity is standard, see for example Ref.~\cite{HS},
and it is equivalent to considering
$u(r=-h/2)=u(r=h/2).$
%
Under this scheme of the spatial discretization, and using zero Dirichlet
boundary conditions at $r=R$ (i.e., fixed edges), we employ the standard
shooting method in time ---implemented by dint of a
fourth-order explicit and symplectic Runge-Kutta-Nystrom method developed in
Ref.~\cite{Calvo}, with time step equal to $T/300$ for the shooting method
and $T/1500$ for the Floquet analysis (see Appendix)---
and seek periodic solutions, for each given temporal period. Then, for each periodic
solution we compute the corresponding stability spectrum via  Floquet
analysis (see Appendix for details).

\begin{figure}[tb]
\begin{center}
\includegraphics[width=7.0cm]{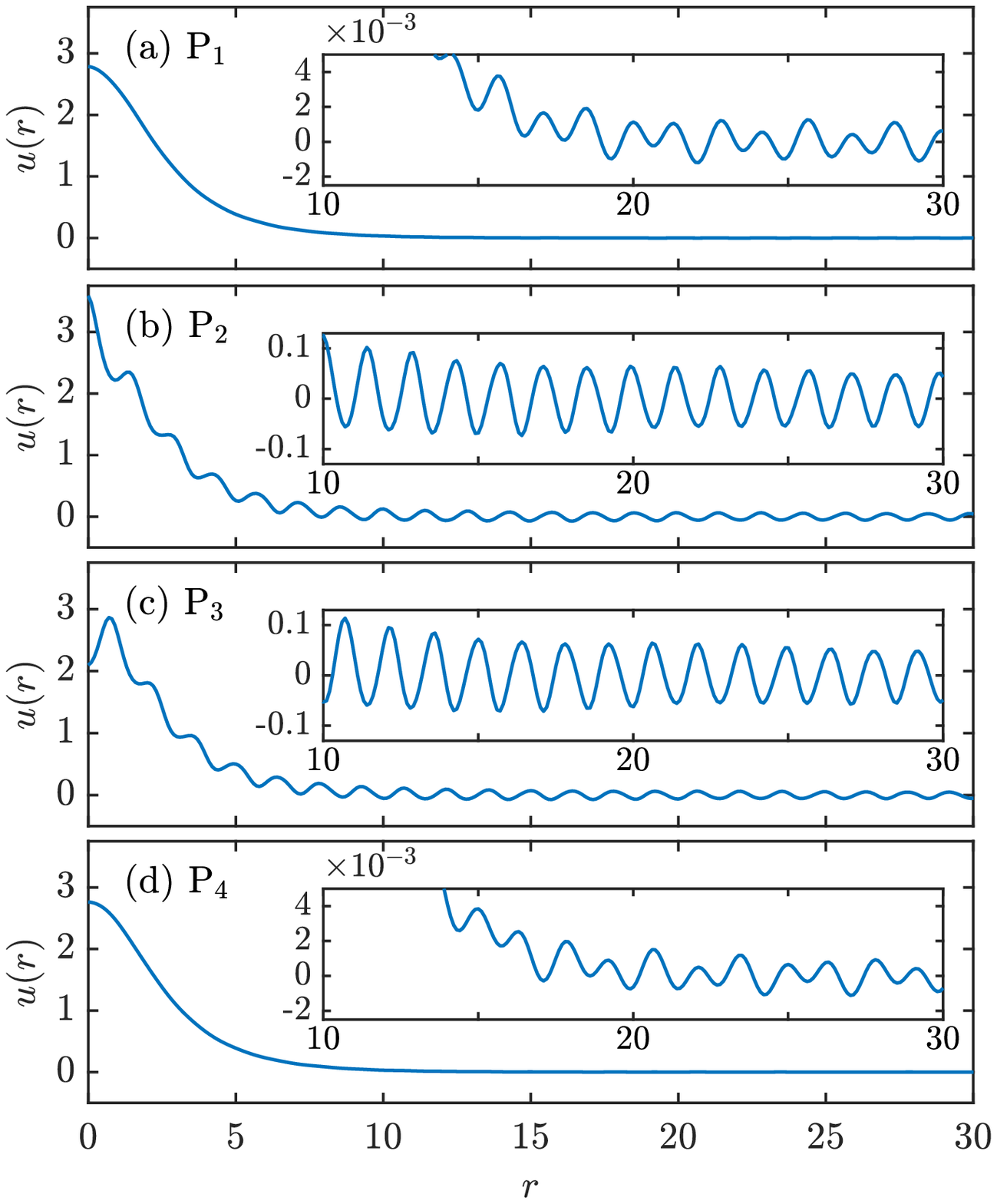}
\end{center}
\caption{(Color online). Radial breather profiles at the points P$_{1}$, P$_{2}$, P$_{3}$,
and P$_{4}$, depicted by large red dots in panel (b) of Fig.~\ref{fig:breather_energy}. }
\label{fig:profilesAB}
\end{figure}

\begin{figure}[htb]
\begin{center}
\includegraphics[width=7.0cm]{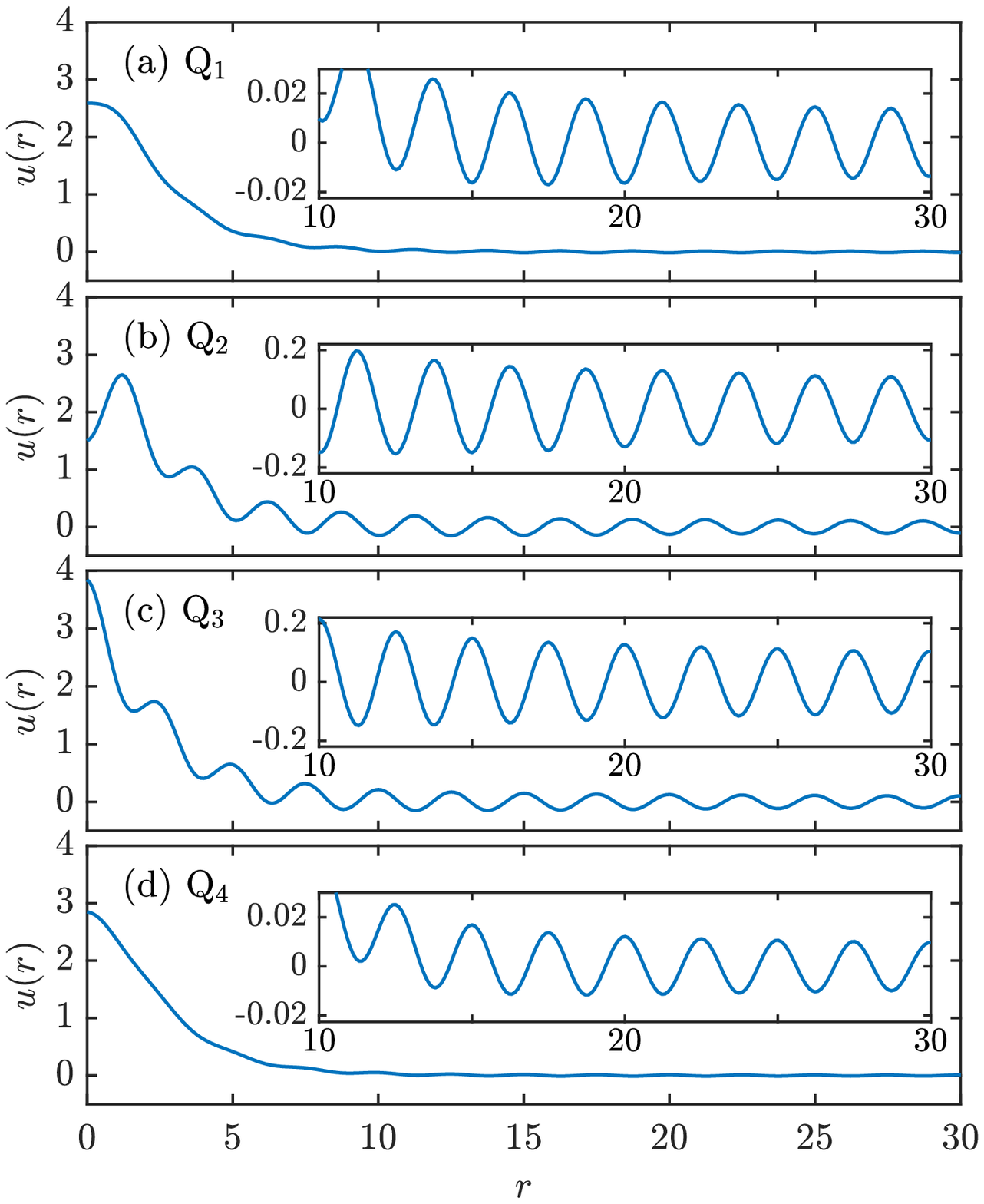}
\end{center}
\caption{(Color online). Radial breather profiles at the points Q$_{1}$, Q$_{2}$, Q$_{3}$,
and Q$_{4}$, depicted by large red dots in panel (c) of Fig.~\ref{fig:breather_energy}. }
\label{fig:profilesBC}
\end{figure}

To gain insight into the stability (and the ensuing dynamics) of radial (intraband)
breathers and their hybridization with phonon modes, we have considered a
small interval around $\omega_{\mathrm{b}}=0.9$.
We have verified that for the cases
under consideration, where $\omega_{\mathrm{b}}$ is close to $1$, the Fourier
coefficients associated with the breathers
decay relatively fast with the increase of their order. This
allows us to resolve the solutions with satisfactory
resolution. Namely, we have corroborated that increasing the temporal
resolution by including more Fourier modes does not lead to any visible
change in the obtained stability spectra.

The main stability results for radial breathers with frequencies close to
$\omega_{\mathrm{b}}=0.9$ are summarized in Fig.~\ref{fig:breather_energy}.
Specifically, panels (a)--(c) depict the total energy [$E=\iint \mathcal{E}\,dxdy$,
with energy density defined as per Eq.~(\ref{eq:energy})] versus
$\omega_{\mathrm{b}}$. Here, three distinct branches (A, B, and C) of the
radial branches are found. In analogy to the 1D phantom breathers of
Ref.~\cite{phantom}, there is branch A to the left, which terminates at a turning
point, $\omega_{\mathrm{b}}=\omega_{A}=0.89990961$, and central branch B
that also finishes at a turning point, $\omega_{\mathrm{b}}=\omega_{B}=0.90215058$.
Herein, we mostly focus on branch B. Panels (b) and (c) depict
zoomed areas from (a) around the frequencies where, respectively, branch A
turns and branch B starts, or branch B turns and branch C starts. These two
bifurcations occur, due to the resonance of the \emph{seventh harmonic} of
the $\omega_{\mathrm{b}}$ with phonons, respectively, at points
$\omega_{\mathrm{phonon}}/7=\omega_{\mathrm{b}}^{A}=0.8999686$
and $\omega_{\mathrm{phonon}}/7=\omega_{\mathrm{b}}^{B}=0.902735$.
These phonon
frequencies produce, via the dispersion relation~(\ref{disp}), the wavenumber
observed in the tails of the breather, confirming that the resonance between
the breather and phonons takes place. The left and central branches
join, via turning points, through additional solution branches
($\mathrm{A}^{\prime }$ and $\mathrm{B}^{\prime }$, respectively,
see Fig.~\ref{fig:breather_energy}).

Note that breathers belonging to branch B-$\mathrm{B}^{\prime }$ have the
amplitude maximum displaced from $r=0$ (that is, the solutions are shaped as
\textit{ring breathers}) at $\omega_{\mathrm{b}}<\omega_{r,A}=0.8999102$
and $\omega_{\mathrm{b}}>\omega_{r,B}=0.902137$. Indeed, Figs.~\ref{fig:profilesAB}
and \ref{fig:profilesBC} display the breather profiles
---again for zero initial velocity $u_t(r,t=0)=0$---
close to the transitions between branches, \textit{viz}., respectively,
$\mathrm{A\rightleftarrows B}$ and $\mathrm{B\rightleftarrows C}$, and,
specifically, in panel (c) of Fig.~\ref{fig:profilesAB} and panel (b) of
Fig.~\ref{fig:profilesBC}, the profiles do not have a maximum at $r=0$. This
qualitative change of the breathers' shape can be understood by noting that,
as mentioned above, the seventh harmonic of these breathers is resonant with
phonon waves, its amplitude being $\simeq 0.1$ times the amplitude of the
fundamental (first) harmonic. As a
consequence, the breather progressively resembles a delocalized phonon wave
when the frequency is decreased. Thus, the rings in the breather shape at
$\omega_{\mathrm{b}}<\omega_{r,A}$ are a consequence of the hybridization
with phonons.
%
%
%
%

\begin{figure}[tb]
\begin{center}
\includegraphics[width=8.0cm]{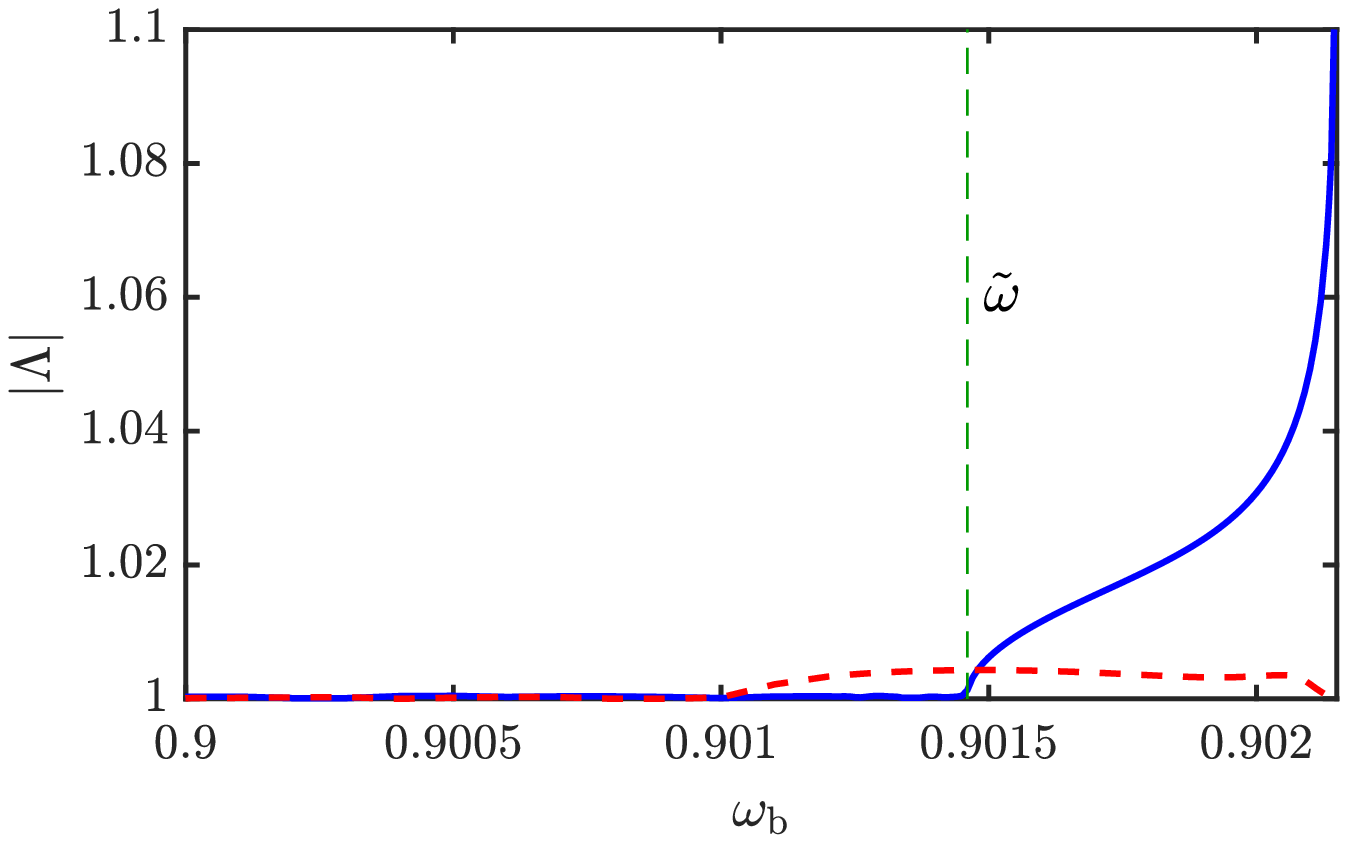}
\end{center}
\vspace{-0.4cm}
\caption{(Color online) Largest Floquet multipliers for $k_\theta=0$
(the solid blue line) and $k_\theta=1$ (the red dashed line) for breathers
belonging to branch B. The strongest instability, which takes place at
$\omega_\mathrm{b}>\tilde{\omega}=0.90146$, is
exponential (monotonously growing),
while other instabilities are oscillatory.}
\label{fig:Floqet_Bbranch}
\end{figure}

Finally, we address the stability of the radial breather solutions. Our
numerical Floquet stability analysis suggests that the breathers belonging
to branch $B$ are mainly affected by radial perturbations, i.e., those with
$k_{\theta}=0$ in Eq.~(\ref{wkya}), and that the only exponential
instabilities that occur are related to the existence of minima in the
energy-vs.-frequency dependence, i.e., ones of the kind predicted in
Ref.~\cite{energy}. Because of this, breathers belonging to branch B in
Fig.~\ref{fig:breather_energy} are subject to exponential instability at
$\omega_{\mathrm{b}}>\tilde{\omega}=0.90146$, where $\tilde{\omega}$ is the
energy-minimum point on B branch, denoted by the vertical green dashed line
in the figure. The full stability spectrum for the radial-breather solutions
in branch B is shown in Fig.~\ref{fig:Floqet_Bbranch}. As expected, the
breather becomes exponentially unstable to radial perturbations (see the
solid blue line representing the FM corresponding to radial
perturbations) at $\omega_{\mathrm{b}}>\tilde{\omega}$. In addition, the
breathers belonging to this branch are also subject to an oscillatory
instability at $\omega_{\mathrm{b}}>0.8999095$, although its magnitude is
small ($|\Lambda |\lesssim 1.005$). We have checked that this weak
oscillatory instability is not a numerical artifact, as changing $k_{m}$
[controlling the number of Fourier modes, per Eq.~(\ref{series})] or
the integration step in the Floquet analysis do not alter the spectrum.
Changing the discretization parameter would also lead to the appearance of
new phonon modes and, at the same time, new resonances that would change the
existence interval of the radial breathers. In any case, these weak
oscillatory instabilities should not have a noticeable effect on the
breathers' dynamics. The oscillatory instabilities were only found for
$k_{\theta }=1$ in Eq.~(\ref{wkya}). The breathers belonging to branch
$\mathrm{B}^{\prime}$ are exponentially unstable against the perturbations
with $k_{\theta}=0$ at all values of $\omega_{\mathrm{b}}$, and feature
weak oscillatory instabilities for $k_{\theta}\geq 1$. A brief exploration
of the stability for breathers belonging to branch A indicates that the
results are very similar to those reported here for branch B.

\section{Conclusions}
\label{sec:conclu}

We have studied the extension of sG breathers to 2D, in the form of quasi-1D
breather stripes and localized radial breathers. Starting from the long-wave
MI (modulational instability) of the stripe, discovered in the seminal work
by Ablowitz and Kodama~\cite{kodama} and confirmed herein,
we have expanded the MI analysis to
arbitrary wavelengths of the transverse perturbations. In the limit of
small-amplitude broad breathers, with frequency $\omega_{\mathrm{b}}\rightarrow 1$,
we have employed the asymptotic multiscale expansion method
to match the dynamics of the breather stripes to that of bright soliton
stripes in the framework of the 2D NLS equation. Perhaps more importantly,
we have developed a novel version of the VA (variational approximation),
whereby the 2D dynamics of the sG breather stripe is reduced to a
filament-type evolution equation for the amplitude of the breather along the
stripe. The VA allowed us to formulate an approximate reduced model that not
only captures the necking MI (i.e., the dependence of the MI gain on the
perturbation wavenumber), but is also able to predict the dynamics beyond
the linear instability setting, and up to close to the breakup of the stripe by
the growing necking perturbations.

It is precisely the necking MI of sG breather stripes that results in the
nucleation of temporally oscillatory, spatially localized ``blobs" in
the form of radial breathers. Numerical results demonstrate that the blobs can
collide or merge with nearby ones. The apparent robustness of these modes
has prompted us to study their existence and stability in more detail. It is
important to note that,  contrary to the 1D sG equation, its 2D
counterpart is not integrable, hence genuine radial breathers cannot exist
in infinite domains for infinitely long times due to resonances of multiple
harmonics of the breather with the phonon (continuous) spectrum.
Nonetheless, in the finite domain it is possible to find intraband
breathers, similarly to how it was done for 1D \textit{phantom breathers}
that were discovered in Ref.~\cite{phantom}. In fact, our results
show that such intraband radial breathers, possessing a
``nanopteron" spatially oscillatory tail, do exist and may be stable in the
finite domain. The full bifurcation structure of such radial breathers with
arbitrary values of the oscillation frequency, $\omega_{\mathrm{b}}$, is
quite complex. Nonetheless, we have performed a detailed numerical study
for the existence of broad small-amplitude radial breathers
for $\omega_{\mathrm{b}}$ close to $1$, in the vicinity of some of the underlying resonances. We
have identified several solution branches and their stability, by means of
numerical Floquet analysis, finding that such breathers may indeed be
stable in the finite domain.

There are numerous avenues deserving further study. For instance, a more
complete characterization of the bifurcation scenarios for arbitrary values
of $\omega_{\mathrm{b}}$ may be able to elucidate further nontrivial bifurcations
and destabilization scenarios. Also, the dynamics and interactions of
radial breathers clearly merit further study: for instance, a
challenging objective is to investigate the outcome of interactions between
radial breathers (including collisions between moving ones), as a function
of their relative phase, velocity, and frequencies. Lastly, it is natural to
generalize the consideration for 3D settings, and examine whether
spherically symmetric radial breathers may survive in a finite domain,
and/or whether they may be produced spontaneously via MI of quasi-2D
breather planes embedded in 3D. These topics are presently under
consideration, and results will be reported elsewhere.

\section*{Acknowledgments}
This material is based upon work supported by the US National Science Foundation
under Grants PHY-1602994 and DMS-1809074 (P.G.K.).
P.G.K.~also acknowledges support from the Leverhulme Trust via a Visiting Fellowship
and thanks the Mathematical Institute of the University of Oxford for its hospitality
during this work.
R.C.G.~gratefully acknowledges support from the US National Science Foundation
under Grant PHY-1603058.
J.C-M. thanks the Regional Government of Andalusia under the project P18-RT-3480 and MICINN, AEI and EU (FEDER program) under the project PID2019-110430GB-C21.
B.A.M.~appreciates support provided by Israel Science Foundation through grant Np.~1286/17.

\section*{Appendix: Numerical implementation for periodic states
and their Floquet stability analysis}
\label{sec:numerical}

In this Appendix, we outline the Floquet analysis that was deployed for quasi-1D
breathers embedded in rectangular domains and radial breathers in circular domains.
These breathers are $T$-periodic solutions corresponding to steady states of the
dynamics after time evolves from $t=0$ to $t=T$. Furthermore, as these solutions
only depend (non-trivially) on a single spatial coordinate ---the $x$- or $r$-direction
for stripe and radial breathers respectively---, they can be numerically computed as
1D solutions along this longitudinal direction.
Nonetheless, it is crucial to note that the stability for these solutions has to be
computed in the {\em full} 2D domain where these breather structures are embedded in.
Namely, perturbations to these solutions need to be followed, not only along the
longitudinal direction, but more importantly, along the transverse direction.
To this end, as described below, we follow the growth of plane-wave {\em transverse}
perturbations with wavenumbers $k_y$ and $k_\theta$, along the $y$- and $\theta$-directions
for, respectively, the breather stripes in the Cartesian plane and radial breathers
in polar coordinates.
Thus, each wavenumber, $k_y$ or $k_\theta$, yields a 1D perturbation
equation determining the stability of the corresponding modulations
over one period $T$ that is treated, as detailed below, using Floquet
analysis to obtain its corresponding eigenvalues (eigenfrequencies)
and eigenvectors (eigenmodes).

The first step in our numerical computations is to discretize in space the solution
and its corresponding derivatives.
To this end, we consider a finite-difference scheme with a (spatial) discretization
parameter $h=0.1$, which transforms the PDE for $u(x,t)$ into a set of
$N$ coupled ODEs for $u_n\equiv u(x_n,t)$ on the discrete grid $\{x_n\}$.
The discretized version of the quasi-1D sG equation (also known as the
Frenkel-Kontorova model) reads:
\begin{equation}
\ddot{u}_{n}+\sin (u_{n})+\frac{1}{h^{2}}(u_{n+1}-2u_{n}+u_{n-1})=0,
\end{equation}
where $n=-N/2\ldots N/2$. Similarly, for the radial breather the
corresponding discretized radial sG equation may be written as
\begin{equation}
\ddot{u}_{n} +\sin (u_{n})+\frac{1}{h^{2}}(u_{n+1}-2u_{n}+u_{n-1})
+\frac{1}{2nh^{2}}(u_{n-1}-u_{n+1})\,,n=1\ldots N
\end{equation}

To produce breathers in the numerical form, we have made use of two
different techniques, based on the fact that the solutions are $T$-periodic,
with $T=2\pi /\omega_{\mathrm{b}}$: (i) a shooting method,
based on the consideration of the map,
\begin{equation}
\mathbf{Y}(0)\rightarrow \mathbf{Y}(T),\qquad \mathbf{Y}(t)=\left[
\begin{array}{c}
\{u_{n}(t)\} \\
\{\dot{u}_{n}(t)\} \\
\end{array}
\right] ,
\end{equation}
and (ii) a Fourier-transform implementation, based on expressing the solution
of the discretized dynamical equations in the form of a truncated Fourier
series:
\begin{equation}
u_{n}(t)=z_{0,n}+2\sum_{k=1}^{k_{m}}z_{k,n}\cos (k\omega_\mathrm{b} t),
\label{series}
\end{equation}
with $k_{m}$ being the maximum of the absolute value of the $k$ in our
truncation of the full Fourier series. In the numerics, $k_{m}=11$ was
chosen. Note that because of the spatial evenness of the sG potential [$V(u)=1-\cos (u)$],
the Fourier coefficients with even $k$ are zero and, consequently, only
odd harmonics of $\omega_{\mathrm{b}}$ can resonate with phonons.

After the introduction of Eq.~(\ref{series}) in the dynamical equations, one
gets a set of $N\times (k_{m}+1)$ nonlinear, coupled algebraic equations.
For a detailed explanation of these methods, the reader is referred to
Refs.~\cite{AMM99,Marin}. In both methods, the continuation in the frequency is
implemented via the path-following (Newton-Raphson) method. The
Fourier-transformed methods have the advantage, among others, of providing
an explicit analytical form of the Jacobian, which makes the calculations
faster. However, for frequencies smaller than $\sim 0.6$, the Fourier
coefficients decay slowly and a large number of coefficients should be kept
to produce an accurate solution, which makes the Jacobian numerically
expensive. Because of this, we developed the analysis based on method (i)
to quasi-1D breathers
in the range $0.1\leq \omega_{\mathrm{b}}<1$, whereas the Fourier-transform
methods (ii) were applied to the radial breathers, when we focused on frequencies
around $\omega_{\mathrm{b}}=0.9$.

To study the spectral stability of the breathers, we introduce a small
perturbation $\xi_{n}(t)$ to a given solution $u_{n,0}(t)$ of the
discretized dynamical equations as $u_{n}(t)=u_{n,0}(t)+\xi_{n}(t)$
~\cite{aubry}. Then, defining the perturbation as $\xi_{n}=\xi_{x,n}\exp (ik_{y}y)$
in the quasi-1D breather stripe case yields
\begin{equation}
\ddot{\xi}_{x,n}+\left[\cos (u_{n})+k_{y}^{2}\right]\xi_{x,n}+\frac{1}{h^{2}}
(\xi_{x,n+1}-2\xi_{x,n} +\xi_{x,n-1})=0.
\notag
\end{equation}
On the other hand, for the radial breather the corresponding dynamics for the
perturbation may be written as
\begin{equation}
\ddot{\zeta}_{r,n} + \left[\cos (u_{n})+\frac{k_{\theta }^{2}}{n^{2}h^{2}}\right] \zeta_{r,n}
+\frac{1}{h^{2}}(\zeta_{r,n+1}-2\zeta_{r,n}+\zeta_{r,n-1})
 +\frac{1}{2nh^{2}}(\zeta_{r,n-1}-\zeta_{r,n+1})=0\,,
\notag
\end{equation}
as $\zeta_{n}=\zeta_{r,n}\exp (ik_{\theta }\theta )$. Due to the (temporal) periodicity
of the solutions, Floquet analysis must be employed. In such a case, the
stability properties are determined by the spectrum of the Floquet operator
$\mathcal{M}$ (whose matrix representation is the monodromy), defined as:
\begin{equation}
\left(
\begin{array}{c}
\{\xi_{n}(T)\} \\
\{\dot{\xi}_{n}(T)\} \\
\end{array}
\right) =\mathcal{M}\left(
\begin{array}{c}
\{\xi_{n}(0)\} \\
\{\dot{\xi}_{n}(0)\} \\
\end{array}
\right) ,  \label{eq:monodromy}
\end{equation}
for the perturbation $\xi_{n}$ in the quasi-1D breather stripe case and an
identical equation for $\zeta_n$ for the radial breather case.
The $2N\times 2N$ monodromy eigenvalues $\Lambda \equiv \exp (i\Theta )$ are
dubbed the FMs (Floquet multipliers), with \emph{Floquet exponents} (FEs)
$\Theta$. A consequence of the fact that the Floquet operator is real is
that, if $\Lambda $ is an FM, $\Lambda^{\ast }$ is an FM
too. Further, because of the symplecticity of the Floquet
operator, $1/\Lambda $ is also an FM. In other words, FMs
always come in quadruplets $(\Lambda ,\Lambda^{\ast },1/\Lambda
,1/\Lambda^{\ast })$ if the monodromy eigenvalues  are complex, and in pairs
$(\Lambda ,1/\Lambda )$ if the eigenvalues are real. As a consequence, a
necessary and sufficient condition for a breather to be linearly stable is
that $\Theta $ must be real (i.e., that the corresponding FMs lie on the unit
circle in the complex plane). Finally, note that FMs $\Lambda $ are
related to eigenvalues $\lambda $ through relation $\Lambda =\exp (\lambda T)$.

\end{document}